\documentclass[12pt]{article}
\usepackage{amsmath,amssymb,amsfonts,graphicx,slashed,color,cite}
\usepackage{color}
\usepackage[colorlinks=true,linkcolor=blue,citecolor=blue,linktoc=all]{hyperref}

\def\leigh{Robert G. Leigh}
\def\uiucaddress{\small Department of Physics, University of Illinois, 1110 W. Green St., 
Urbana IL 61801-3080, U.S.A. }
\def\title{\LARGE{Entanglement Entropy \&\ Anomaly Inflow
}}

\parskip=\baselineskip

\newcommand{\myfig}[3]{
	\begin{figure}[ht]
	\centering
	\includegraphics[width=#2cm]{#1}\caption{#3}\label{fig:#1}
	\end{figure}
	}

\newcommand{\cDsl}{{{\cal D}\kern-.65em /\,}}
\newcommand{\cDslsm}{{{\cal D}\kern-.5em /\,}}
\newcommand{\nabslsm}{\nabla\kern-.55em /}
\newcommand{\pasl}{\pa\kern-.55em /}
\newcommand{\psl}{p\kern-.45em /}
\newcommand{\Dsl}{D\kern-.65em /}
\newcommand{\Asl}{A\kern-.55em /}
\newcommand{\nabsl}{\nabla\kern-.65em /\kern+.2em}
\newcommand{\qsl}{q\kern-.5em /}
\newcommand{\ksl}{k\kern-.5em /}
\newcommand{\rsl}{r\kern-.5em /}

\newcommand{\ue}{\underline{e}}

\newcommand{\cDslLCsq}{{\stackrel{\circ}{\cDsl^{\kern2pt 2}}}}

\newcommand\cc[1]{#1^{^{\kern-6pt \circ}}\kern2pt}

\newcommand{\re}{\mathbb{R}}

\newcommand{\pa}{\partial}

\newcommand{\beq}{\begin{equation}}
\newcommand{\eeq}{\end{equation}}
\newcommand{\beqn}{\begin{eqnarray}}
\newcommand{\eeqn}{\end{eqnarray}}

\newcommand{\bdx}{\mathbf{x}}

\newcommand{\hT}{\widehat{T}}
\newcommand{\hK}{\widehat{K}}

\newcommand{\Bx}{\boldsymbol{x}}

\def\dalemb#1#2{{\vbox{\hrule height .#2pt
\hbox{\vrule width.#2pt height#1pt \kern#1pt
\vrule width.#2pt}
\hrule height.#2pt}}}

\begin{document}

\begin{center}
\title
\end{center}
\vskip 2 cm
\centerline{{\bf 
Taylor L. Hughes, \leigh, Onkar Parrikar \&\ Srinidhi T. Ramamurthy}}

\vspace{.5cm}
\centerline{\it \uiucaddress}
\vspace{2cm}

\begin{abstract}
\noindent We study entanglement entropy for parity-violating (time-reversal breaking) quantum field theories on $\re^{1,2}$ in the presence of a domain wall between two distinct parity-odd phases. The domain wall hosts a 1+1-dimensional conformal field theory (CFT) with non-trivial chiral central charge. Such a CFT possesses gravitational anomalies. It has been shown recently that, as a consequence, its intrinsic entanglement entropy is sensitive to Lorentz boosts around the entangling surface. Here, we show using various methods that the entanglement entropy of the three-dimensional bulk theory is also sensitive to such boosts owing to parity-violating effects, and that the bulk response to a Lorentz boost precisely cancels the contribution coming from the domain wall CFT. We argue that this can naturally be interpreted as \emph{entanglement inflow} (i.e., inflow of entanglement entropy analogous to the familiar Callan-Harvey effect) between the bulk and the domain-wall, mediated by the low-lying states in the entanglement spectrum. These results can be generally applied to 2+1-d topological phases of matter that have edge theories with gravitational anomalies, and provide a precise connection between the gravitational anomaly of the physical edge theory and the low-lying spectrum of the entanglement Hamiltonian.
\end{abstract}

\pagebreak
%
%
\section{Introduction and Preliminaries}
In recent years, entanglement entropy has acquired a central position in the study of quantum field theories. It has emerged as a powerful tool to probe the structure of quantum states primarily because: (i) it is sufficiently non-local to capture certain global properties, and (ii) it is geometric by definition and hence universal in its applicability. As a result, entanglement entropy has provided great insights in a wide class of systems such as relativistic field theories \cite{Calabrese:2004eu,1751-8121-42-50-504005,Casini:2009sr}, 2+1-d topologically ordered phases of matter \cite{hammakitaev,Kitaev:2005dm,PhysRevLett.96.110405,Dong:2008ft,lihaldane,Flammia2009,Regnault2009,Yao2010,sterdyniak2011a,hermanns2011}, strongly-coupled theories with holographic duals, \cite{Ryu:2006bv,Ryu:2006ef} etc. There have also been suggestions that entanglement might play a crucial role in understanding the emergence of geometry in the AdS/CFT correspondence \cite{VanRaamsdonk:2010pw,Faulkner:2013ica}.

Entanglement entropy is defined as follows -- consider a density matrix $\rho$ corresponding to a pure state defined on a Cauchy surface $\Sigma$. Let us partition $\Sigma$ into two subregions $A$ and $\bar{A}$. For local quantum field theories, we expect the Hilbert space $\mathcal{H}_{\Sigma}$ to factorize into the tensor product $\mathcal{H}_{\Sigma}=\mathcal{H}_{A}\otimes \mathcal{H}_{\bar{A}}$. If this is the case, we can trace over $\mathcal{H}_{\bar{A}}$ to obtain the reduced density matrix
\beq
\rho_A = \mathrm{Tr}_{\mathcal{H}_{\bar{A}}}(\rho)
\eeq 
which contains all the relevant information pertaining to the subregion $A$. Then the \emph{entanglement entropy}  between $A$ and $\bar{A}$ is defined as the von Neumann entropy of $\rho_A$
\beq
S_{EE}[\rho_A] = -\mathrm{Tr}_{\mathcal{H}_A}\left(\rho_A\mathrm{ln}\;\rho_A\right).
\eeq
In this context, the boundary $\pa A$ of $A$ is referred to as the \emph{entangling surface} (or entanglement cut). It is also useful to define the \emph{entanglement Hamiltonian} (also known as the modular Hamiltonian) $\widehat{H}_E$ in terms of $\rho_A$ as
\beq
\rho_A \equiv e^{-\widehat{H}_E}.
\eeq
Entanglement entropy satisfies a number of important properties. One such property which will be of relevance to us here, is that entanglement entropy is actually a function of the \emph{domain of dependence} $D[A]$ (see Fig.~\ref{fig:fig1.pdf}).\footnote{The domain of dependence $D[A]$ of the region $A$ is the set of all points $p$ for which every inextendible causal curve through $p$ intersects $A$.} This means that if we were to pick a different Cauchy surface $\Sigma'$ and correspondingly a different subregion $A'$ such that $D[A]=D[A']$, then the entanglement entropy $S_{EE}[A]$ would be equal to $S_{EE}[A']$ \cite{Headrick:2014cta}. Of course, the entanglement entropy in the vacuum state of a quantum field theory typically diverges, and the above property is really satisfied by a suitably regulated version of the entanglement entropy. 

\myfig{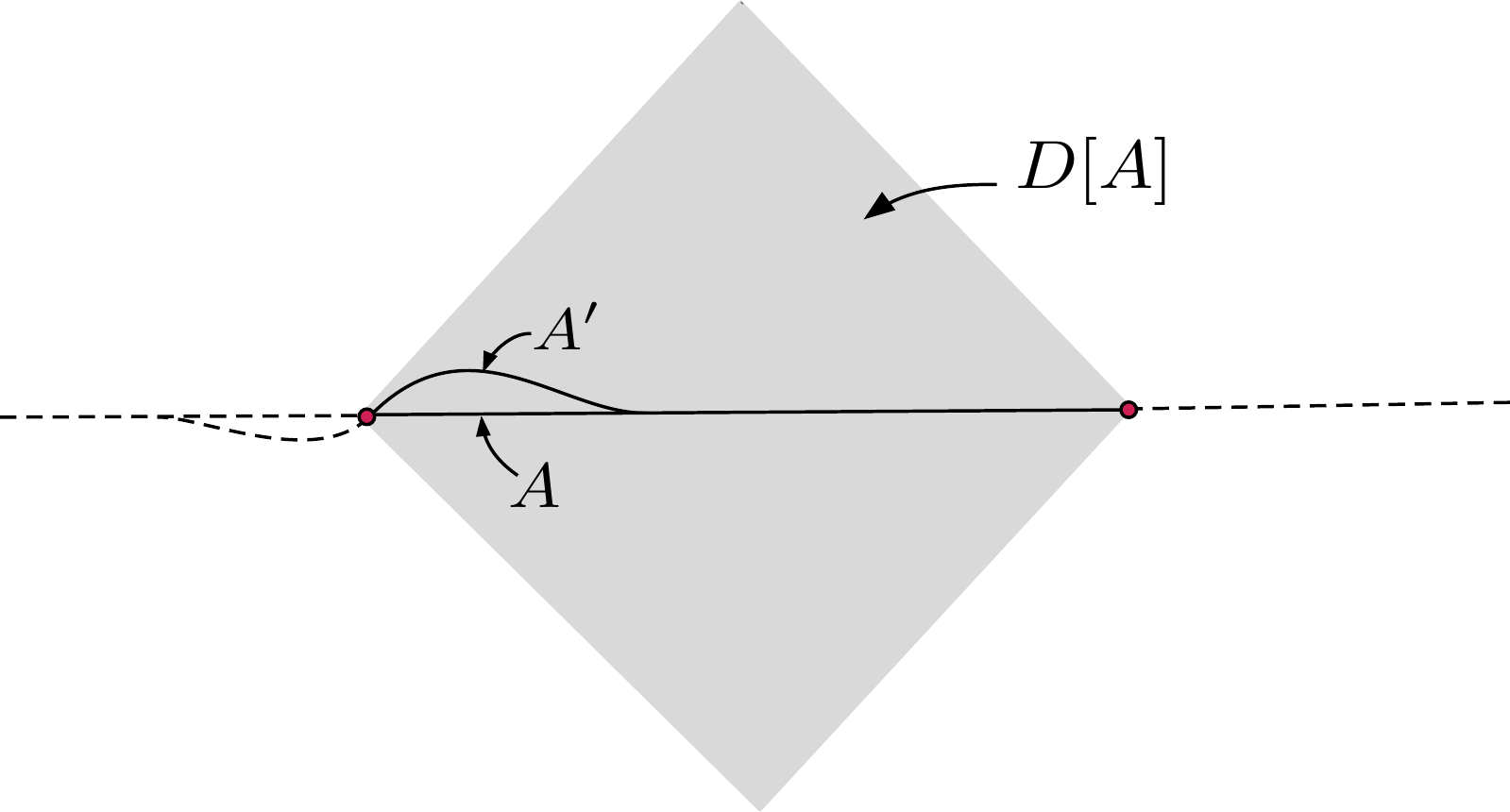}{10}{\small{\textsf{The domain of dependence $D[A]$ of the subregion $A$ (bold line) is the shaded region. If we change $A$ to $A'$ in such a way that $D[A]=D[A']$, then the entanglement entropy remains unchanged. The red dots denote the entangling surface $\pa A$.}}}

In this paper, we are interested in the entanglement entropy in the vacuum state of a parity-violating (equivalently time-reversal breaking) quantum field theory on three-dimensional Minkowski spacetime $\re^{1,2}$, where we pick the region $A$ to be a half-space. By a parity-violating quantum field theory, we mean a quantum field theory which when coupled to background gauge or gravitational fields, contains parity-odd terms in its low-energy effective action. The quintessential examples of such terms, which actually suffice for our purposes, are given by the Chern-Simons terms
\beq \label{CSterms}
S_{eff}[A,g] = \frac{\sigma}{2}\int \mathrm{tr}\;\left(A\wedge dA+\frac{2}{3}A\wedge A\wedge A\right) +\frac{\kappa}{2}\int \mathrm{tr}\;\left(\Gamma\wedge d\Gamma+\frac{2}{3}\Gamma\wedge \Gamma\wedge \Gamma\right)+\cdots
\eeq
where $A$ is a background gauge field, $g$ is a background metric and $\Gamma$ the corresponding Christoffel connection of $g.$ The set of coefficients $\sigma, \kappa$, etc.,  give a characterization of the \emph{phase} of the theory with respect to parity. In this paper, we will mostly be interested in the gravitational Chern-Simons term, and consequently we will restrict our attention only to $\kappa$. 

An interesting scenario in such theories is as follows: if $x$ labels one of the spatial coordinates, then consider a situation where the value of the Chern-Simons coefficient $\kappa$ jumps from $\kappa_-$ to $\kappa_+$ across the codimension-one surface $x=0$. We will refer to this configuration, and in particular the surface $x=0$, as a \emph{parity domain wall}. Such a domain wall typically hosts a 1+1-dimensional conformal field theory (CFT), with a non-trivial chiral central charge $c_L-c_R$. However, it is well known that CFTs with $c_L-c_R\neq 0$ have gravitational anomalies -- this means that when such a CFT is coupled to a background metric, its stress tensor is not conserved \cite{AlvarezGaume:1983ig,AlvarezGaume:1984dr}
\beq\label{consanomaly}
\nabla^{i}\left\langle\hT^{CFT}_{ij}\right\rangle= \frac{c_L-c_R}{96\pi}\epsilon^{kl}\pa_{k}\pa_{m}{\Gamma^m}_{lj}
\eeq
and thus (background) diffeomorphism invariance\footnote{By background diffeomorphism invariance, we mean that the partition function $Z[g_{\mu\nu}]$ regarded as a functional of the background metric, satisfies the Ward identity $Z[g_{\mu\nu}]=Z[g_{\mu\nu}+2\pa_{(\mu}\xi_{\nu)}]$.} is broken. Here $\hT^{CFT}_{ij}$ is the \emph{consistent} stress tensor which is obtained by directly varying the CFT partition function with respect to the metric, and its non-conservation is called the \emph{consistent anomaly}. For the parity domain wall, the anomaly is remedied by the fact that the bulk theory is also not diffeomorphism invariant (in the presence of a domain wall). Indeed, the bulk gravitational Chern-Simons term transforms under diffeomorphisms to precisely compensate for the CFT violation of diffeomorphism invariance, provided that
\beq\label{BBC}
\kappa_+-\kappa_- = \frac{c_L-c_R}{48\pi}.
\eeq
In this case, the full theory, i.e., the bulk theory plus domain wall CFT, respects background diffeomorphism symmetry. This cancellation of the CFT gravitational anomaly by the bulk is an example of Callan-Harvey anomaly inflow \cite{Callan:1984sa}. 

Although the above discussion has been somewhat formal, a simple yet concrete example to keep in mind is a Dirac fermion on $\re^{1,2}$ with a mass domain wall, i.e., the mass of the fermion is a function $m(x) = m_0\varphi(x)$ of one of the spatial coordinates $x$, where $m_0$ is a positive constant and $\varphi(x)$ smoothly interpolates between $-1$ and $+1$ as $x$ goes from $-\infty$ to $+\infty$. It is easy to show by solving the Dirac equation that the domain wall traps a left-moving chiral (Weyl) fermion in the limit $m_0\to \infty$, which can be represented by a free CFT on the domain wall with $c_L-c_R=1$. On the other hand, far from the domain wall on either side, one can integrate out the bulk fermion (which recall becomes very massive as $m_0 \to \infty$) to obtain the bulk effective action. Indeed, one finds that this effective action has a gravitational Chern-Simons term, where the coefficient precisely satisfies equation \eqref{BBC} \cite{AlvarezGaume:1984nf,Parrikar:2014usa}. In fact, this is essentially the macroscopic physics of a condensed matter system dubbed a Chern insulator\cite{haldane1988} which exhibits an integer quantum hall effect (without an external magnetic field) due to the time-reversal/parity-violating Dirac mass. In that context, the gapped bulk region $x<0$ is taken to be the interior of the non-trivial topological Chern insulator, and  $x>0$ is the trivial insulating vacuum. The chiral fermion states on the domain wall are the conducting edge states. Another class of condensed matter systems that exhibit the above type of anomaly inflow physics are fractional quantum hall states -- here the low-energy effective field theory describing the bulk is a topological field theory (usually Chern-Simons theory), while the domain wall CFT is generically a WZW model \cite{wenreview}. In these systems, the gravitational Chern-Simons term in the bulk effective action is generated by the framing anomaly \cite{Witten:1988hf}. 

Let us now consider the entanglement entropy for the class of systems discussed above. Let $\Bx^{\mu} = (t,y,x)$ be coordinates on $\re^{1,2}$, where $t$ parametrizes time. We take the domain wall to lie at $x=0$. We subdivide the spatial slice $\Sigma$ at $t=0$ into two half-spaces $A= \{\Bx \in \Sigma | y>0\}$, and $\bar{A} = \{\Bx \in \Sigma | y<0\}$, with the surface $\pa A$ at $y=0$ being the entangling surface (see Fig.~\ref{fig:fig2.pdf}). 

\myfig{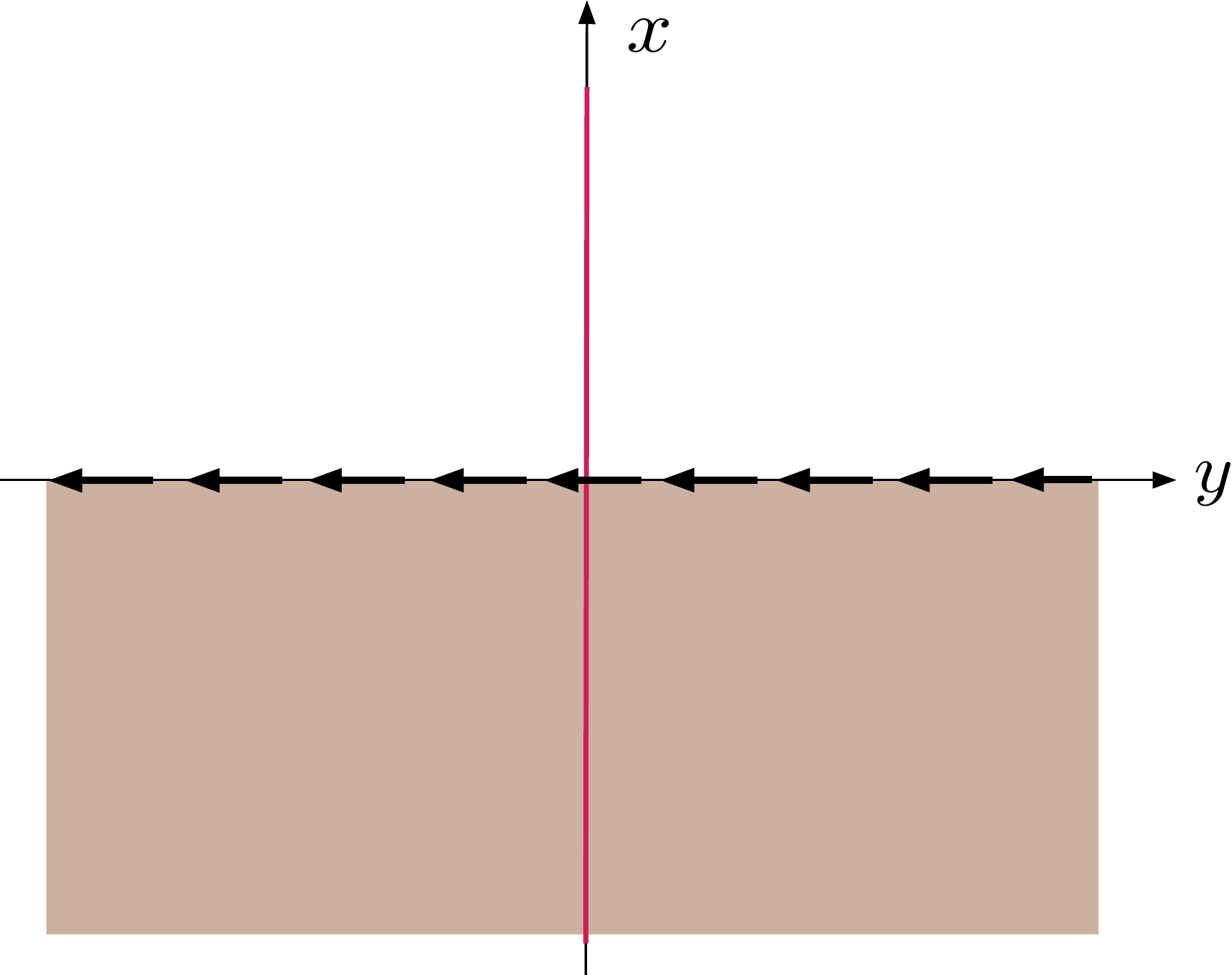}{8}{\small{\textsf{The spatial geometry at $t=0$: the darker region $x<0$ is a particular parity-odd phase with $\kappa=\kappa_-$, while the lighter region $x>0$ is a different phase with $\kappa=\kappa_+$. The domain wall $x=0$ hosts chiral edge states, denoted by bold arrows. The red line $y=0$ is the entangling surface, with $A$ being the half space $y>0$.}}}

In principle one could consider more general entangling surfaces, but we will focus our attention on this case because it simplifies the analysis significantly, and will allow us to make fairly general statements. Now, we are interested in studying how the entanglement entropy between $A$ and $\bar{A}$ depends on an infinitesimal Lorentz boost localized around the entangling surface with the rapidity $\eta$ (see Fig.~\ref{fig:fig3.pdf}). 

\myfig{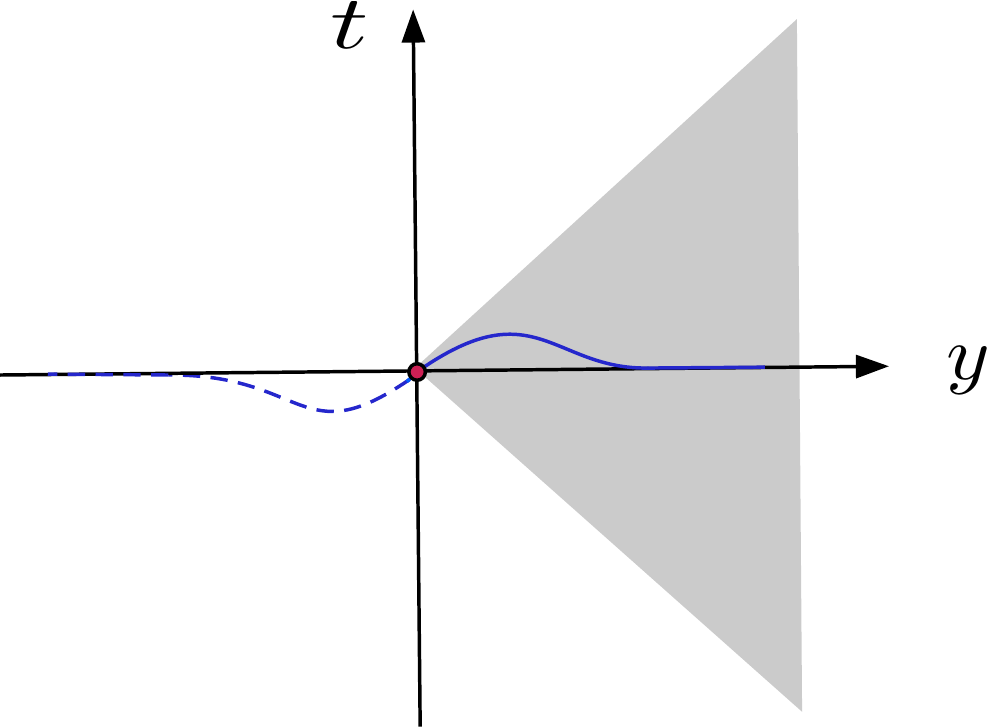}{8}{\small{\textsf{The boosted Cauchy slice (blue) in the $(t,y)$ plane; the $x$ direction is implicit. Since the domain of dependence of $A$ (shaded region) remains unchanged, the entanglement entropy is expected to remain unchanged.}}}

Based on our previous discussion, one expects that the entanglement entropy, being a function of the domain of dependence of $A$, remains invariant under such a boost (which preserves $D[A]$). However, there is a problem here because the entangling surface cuts through the domain wall, and the domain wall CFT has a gravitational anomaly. Indeed, it was pointed out in \cite{Perlmutter:2013paa, Castro:2014tta, Walltalk, IqbalWall, Nishiokatalk, Nishioka:2015uka} (and anticipated previously in \cite{Wall:2011kb}) that the entanglement entropy of a CFT with non-trivial chiral central charge is sensitive to such a local Lorentz boost, and changes by the amount\footnote{References \cite{Perlmutter:2013paa, Castro:2014tta} use a different set-up involving boosted intervals (instead of boosting locally around the entangling surface), and find $\delta S_{EE}=-\eta\frac{c_L-c_R}{6}$. References \cite{Walltalk, IqbalWall} find $\delta S_{EE}$ for a semi-infinite interval to be twice as large as \eqref{DWanomaly}, namely $\delta S_{EE}=-\eta\frac{c_L-c_R}{12}$, which might be akin to a covariant form of the anomaly. Our result agrees with that of \cite{Nishiokatalk, Nishioka:2015uka}. See \cite{IqbalWall,Nishioka:2015uka} for a more detailed comparison.}
\beq\label{DWanomaly}
\delta S_{EE,CFT} = -\eta\frac{c_L-c_R}{24}.
\eeq
Thus, for our expectation of invariance of entanglement entropy under transformations which preserve $D[A]$ to be true, the entanglement entropy of the \emph{bulk} must also change under the local boost by an amount which precisely cancels the CFT contribution. The primary goal of the present paper is to show that this is indeed the case. We will establish this in a number of ways. In Section \ref{sec2.1}, we will derive the response of the bulk entanglement entropy to local boosts by using perturbative methods developed in \cite{Rosenhaus:2014woa, Rosenhaus:2014zza}. In Section \ref{sec2.2}, we will derive the same result using the replica trick because this will give us an understanding of the higher R\'{e}nyi entropies as well. Then in Section \ref{sec2.3}, we study the system by conformally mapping the Rindler wedge $D[A]$ to $\re \times \mathbb{H}^2$ (where $\mathbb{H}^2$ is the two-dimensional hyperbolic space). This will give us a way to understand the cancellation between the bulk and the domain wall as an \emph{inflow} of entanglement entropy mediated by the low-lying states in the entanglement spectrum. Finally, in Section \ref{sec2.4} we will present numerical data which demonstrates the above physics in the specific case of a lattice-regularized Chern-insulator model. 

\section{The bulk entanglement entropy}
Let us now delve into the calculation of the change in the bulk entanglement entropy in response to a local Lorentz boost around the entangling surface. More precisely, we mean the change in the entanglement entropy coming purely from the three dimensional parity-violating bulk theory. We will see that the gravitational Chern-Simons term plays a crucial role in this calculation. 
\subsection{Perturbative method}\label{sec2.1}
In this section, we will use the perturbative techniques developed in \cite{Rosenhaus:2014woa, Rosenhaus:2014zza, Allais:2014ata, Faulkner:2014jva}. Recall that the region $A$ we're interested in is the half-space $y>0$, and so the entangling surface is given by the line $y=0$ (see Fig.~\ref{fig:fig2.pdf}). We start by performing a local boost in the $y$-direction (see Fig.~\ref{fig:fig3.pdf}) given by the following diffeomorphism 
\beq
\label{eq:boostdiff}
\xi(\Bx)= \eta \phi(\Bx) \left(y\pa_{t}+t\pa_y\right)
\eeq
where $\Bx^{\mu} =(t,y,x)$ are coordinates on the full spacetime $\re^{1,2}$. Here $\eta$ is a constant equal to the local rapidity (or the hyperbolic boost angle) close to the entangling surface, and $\phi(\Bx)$ is a smooth function such that $\phi(\Bx)=1$ inside the ball $B_R$ of radius $R$ around $\Bx=0$ (with $R$ much larger than the thickness of the domain wall, or any short distance cutoffs in the theory), and $\phi(\Bx)\to 0$ smoothly outside $B_R$. Note that $\phi$ approaches zero at infinity in \emph{all} directions -- this will simplify our calculations by letting us drop boundary terms at infinity. Before moving on, it is useful to Wick rotate to Euclidean signature by defining $t = i\tau_E$. In this case, the above diffeomorphism becomes
\beq\label{Ediff}
\xi(\Bx)=\eta_E\phi(\Bx)\left(y\pa_{\tau_E}-\tau_E\pa_y\right)
\eeq
where $\eta_E = -i\eta$. 

Now, the above diffeomorphism preserves the entangling surface, but changes the metric around it
\beq
\delta_{\xi}g_{\mu\nu} =2\pa_{(\mu}\xi_{\nu)}
\eeq
where the Greek indices $\mu, \nu$ now run over the bulk Euclidean indices. The change in entanglement entropy to first order in the above perturbation is easy to compute and one finds (see Appendix \ref{appA} for more details)
\beq\label{pert1}
\delta_{\xi} S_{EE,bulk} = -\lim_{a\to 0}\int_{\tau_E^2+y^2\geq a^2} d^3\Bx\;\pa_{(\mu}\xi_{\nu)}(\Bx)\;\left\langle \hT^{\mu\nu}(\Bx)\widehat{H}_E\right\rangle_{\mathrm{conn}.}
\eeq
where $\widehat{H}_{E}$ is the entanglement Hamiltonian, which can be written in terms of the following integral over the region $A$ at $\tau_E=0$
\beq\label{EHmain}
\widehat{H}_E = 2\pi\int_{-\infty}^{\infty}dx'\int_0^{\infty} dy'\;y'\;\hT^{00}(0,y',x')+\mathrm{constant}.
\eeq
The correlation function appearing above is the \emph{connected} Euclidean correlation function on $\re^3$, and so we may drop the constant term in \eqref{EHmain}. To regulate this calculation, we have cut out a tubular neighborhood of radius $a$ around the entangling surface by restricting the integration over $\Bx$ in equation \eqref{pert1} to the region $\tau_E^2+y^2\geq a^2$ (Fig.~\ref{fig:fig4.pdf}); we will send $a\to 0$ at the end of the calculation. Integrating by parts in $\Bx$, we find
\beq\label{pert2}
\delta_{\xi}S_{EE,bulk} =\lim_{a\to 0}\;2\pi a\oint_{\tau_E^2+y^2=a^2}d^2\Bx\int_{-\infty}^{\infty}dx'\int_0^{\infty}dy'\;y'\;\;\xi^{\mu}(\Bx)n^{\nu}(\Bx)\;\left\langle \hT_{\mu\nu}(\Bx)\hT_{00}(0,y',x')\right\rangle
\eeq
where $n^{\mu}(\Bx)$ is the outward pointing unit-normal to the cylinder $\tau_E^2+y^2=a^2$. We have dropped the term proportional to the divergence of the stress-tensor using the Ward identity
\beqn
\pa^{\mu}\left\langle \hT_{\mu\nu}(\Bx)\hT_{\lambda\sigma}(\Bx')\right\rangle &=& \pa_{\nu}\delta^3(\Bx-\Bx')\left\langle \hT_{\lambda\sigma}(\Bx)\right\rangle+2\pa_{(\lambda}\left(\delta^3(\Bx-\Bx')\left\langle\hT_{\sigma)\nu}(\Bx)\right\rangle\right)\nonumber\\
&-&\pa^{\kappa}\delta(\Bx-\Bx')\left\langle\hT_{\kappa\nu}(\Bx)\right\rangle\delta_{\lambda\sigma},
\eeqn
and the fact that the one-point function of the stress tensor has no parity-odd contributions in flat space. We have also dropped all other boundary terms because $\xi(\Bx)\to 0$ far away from $\Bx=0$. Therefore, equation \eqref{pert2} suggests that the change in entanglement entropy can be computed from the two-point function of stress tensors.

\myfig{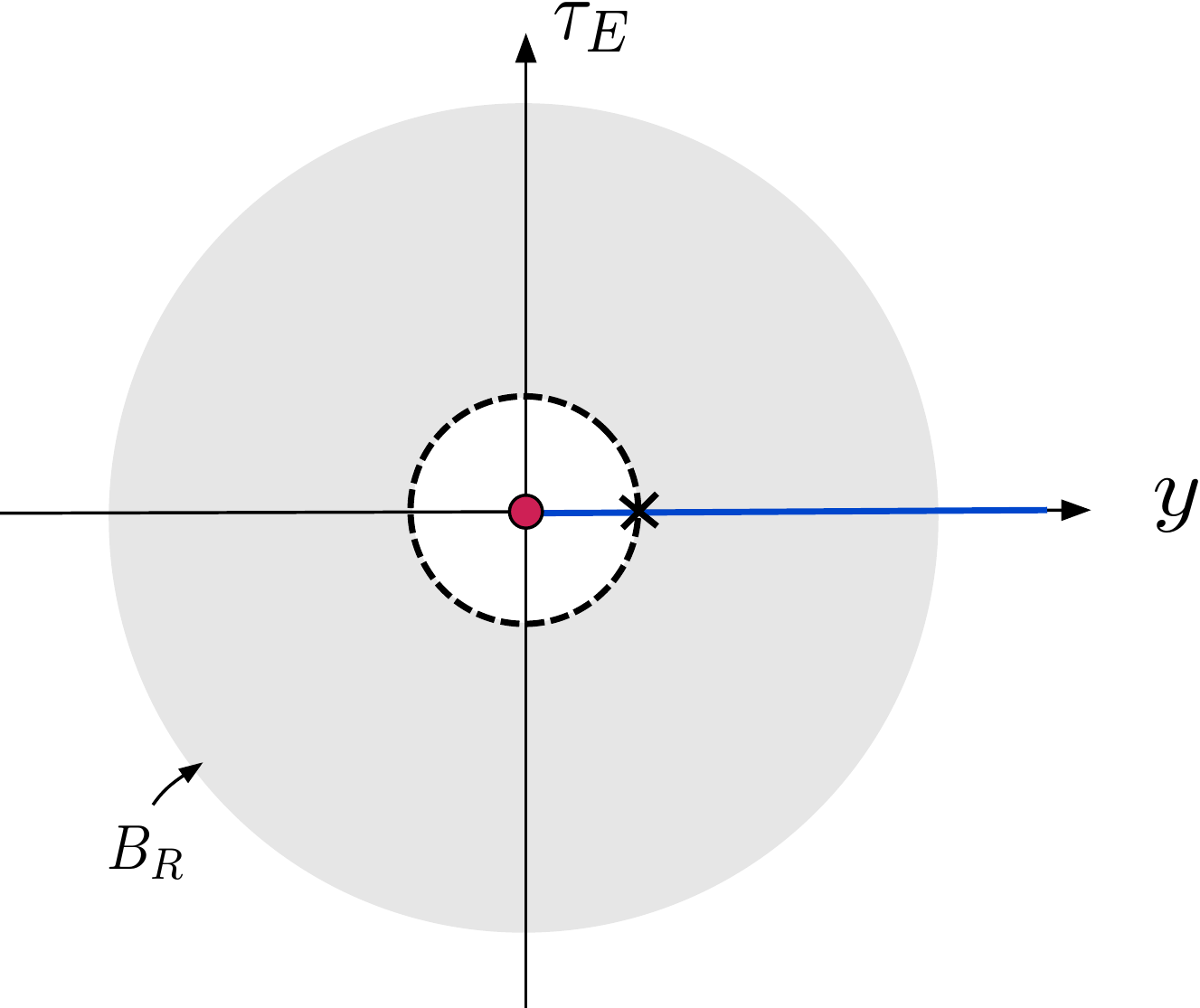}{7}{\small{\textsf{A tubular neighborhood of the entangling surface (red dot) is cut out to regulate the calculation (the $x$ direction is implicitly out of the page). The dotted line is the cut-off surface $\tau_E^2+y^2=a^2$. The entanglement Hamiltonian has support over the blue line. For contact terms in the 2-point function, the integral in equation \eqref{pert2} has support only at the point marked with a cross. The shaded region is the ball $B_R$ where $\xi$ is supported.}}}

Indeed, in theories where parity/time-reversal is violated, there is a parity-odd piece in this two-point function \cite{Leigh:2003ez, Closset:2012vp}
\beq
\left\langle\hT_{\mu\nu}(\Bx)\hT_{\lambda\rho}(\Bx')\right\rangle_{odd}=-2i\kappa\; \epsilon_{(\mu(\lambda\sigma}\left(\pa_{\nu)}\pa_{\rho)}-\delta_{\nu)\rho)}\pa^2\right)\pa^{\sigma}\delta^3(\Bx-\Bx')
\eeq
where the coefficient $\kappa$ is precisely the coefficient of the gravitational Chern-Simons term defined in \eqref{CSterms}.\footnote{\label{HVfootnote}There is also a term lower order in derivatives of the form \cite{witten2007,hughes2011torsional,Hughes:2012vg,parrikar2014}
$$\left\langle\hT_{\mu\nu}(\Bx)\hT_{\lambda\rho}(\Bx')\right\rangle_{odd}= i\zeta\;\epsilon_{(\mu(\lambda\sigma}\delta_{\nu)\rho)}\pa^{\sigma}\delta^3(\Bx-\Bx')+\cdots$$
where $\zeta$ is related to the Hall viscosity response\cite{avron1995,read2009,hughes2011torsional}. It can be checked explicitly that this term does not contribute to the integrals in equation \eqref{pert2}.} However, recall that in the presence of a domain wall at $x=0$, $\kappa$ is \emph{not} a constant but a function $\kappa(x)$ which jumps across the domain wall, i.e., $\kappa(x>>\frac{\ell}{2})=\kappa_+$ and $\kappa(x<<-\frac{\ell}{2})=\kappa_-$ (where $\ell$ is the thickness of the domain wall). The diffeomorphism invariance of the full theory (bulk plus domain wall) imposes the constraint \eqref{BBC}. Taking this into account, we find that the two-point function of stress tensors has additional contributions\footnote{These can be obtained by promoting $\kappa$ to a function $\kappa(\Bx)$ and varying the gravitational Chern-Simons action twice with respect to the metric.}
\beq
\left\langle\hT_{\mu\nu}(\Bx)\hT_{\lambda\rho}(\Bx')\right\rangle_{odd}=\Big(-2i\kappa(x)\pa^{\sigma}+i\pa^{\sigma}\kappa(x)\Big) \epsilon_{(\mu(\lambda\sigma}\left(\pa_{\nu)}\pa_{\rho)}-\delta_{\nu)\rho)}\pa^2\right)\delta^3(\Bx-\Bx')+\cdots
\eeq
where $\cdots$ represents further terms which are not important for the present calculation. Now we simply substitute this expression into the formula \eqref{pert2} for $\delta_{\xi}S_{EE,bulk}$, and compute the integrals. Using the explicit form of $\xi$ given in \eqref{Ediff}, we find 
\beq
\delta_{\xi}S_{EE,bulk} = 2\pi i\eta_E \Big(\kappa_+-\kappa_-\Big)=\eta\frac{(c_L-c_R)}{24}
\eeq
where in the second equality we have reverted back to Lorentzian signature, and used equation \eqref{BBC}. Comparing with equation \eqref{DWanomaly}, we see that the change in the bulk entanglement entropy upon performing a local boost precisely cancels the contribution coming from the domain wall CFT.

\subsection{Replica trick}\label{sec2.2}
To demonstrate the robustness of the above result, let us now repeat this calculation using the replica trick. This also has the added advantage that it will give us an understanding of what is happening with higher R\'{e}nyi entropies. Recall that the $n$th \emph{R\'{e}nyi entropy} is defined by
\beq
S_{n}=\frac{1}{1-n}\mathrm{ln}\;\mathrm{Tr}_{\mathcal{H}_A}\rho_{A}^n. 
\eeq
Note that taking the limit $n\to 1$, $S_n$ approaches the conventional von Neumann entanglement entropy. For integer $n$, there is an explicit prescription to compute these entropies using path-integral techniques\cite{Calabrese:2004eu, Ryu:2006ef}. One introduces the replica manifold $M_n$, which is an $n$-sheeted covering of the original (Euclidean) spacetime manifold branched along the region $A$ (see Fig.~\ref{fig:fig5.pdf}).

\myfig{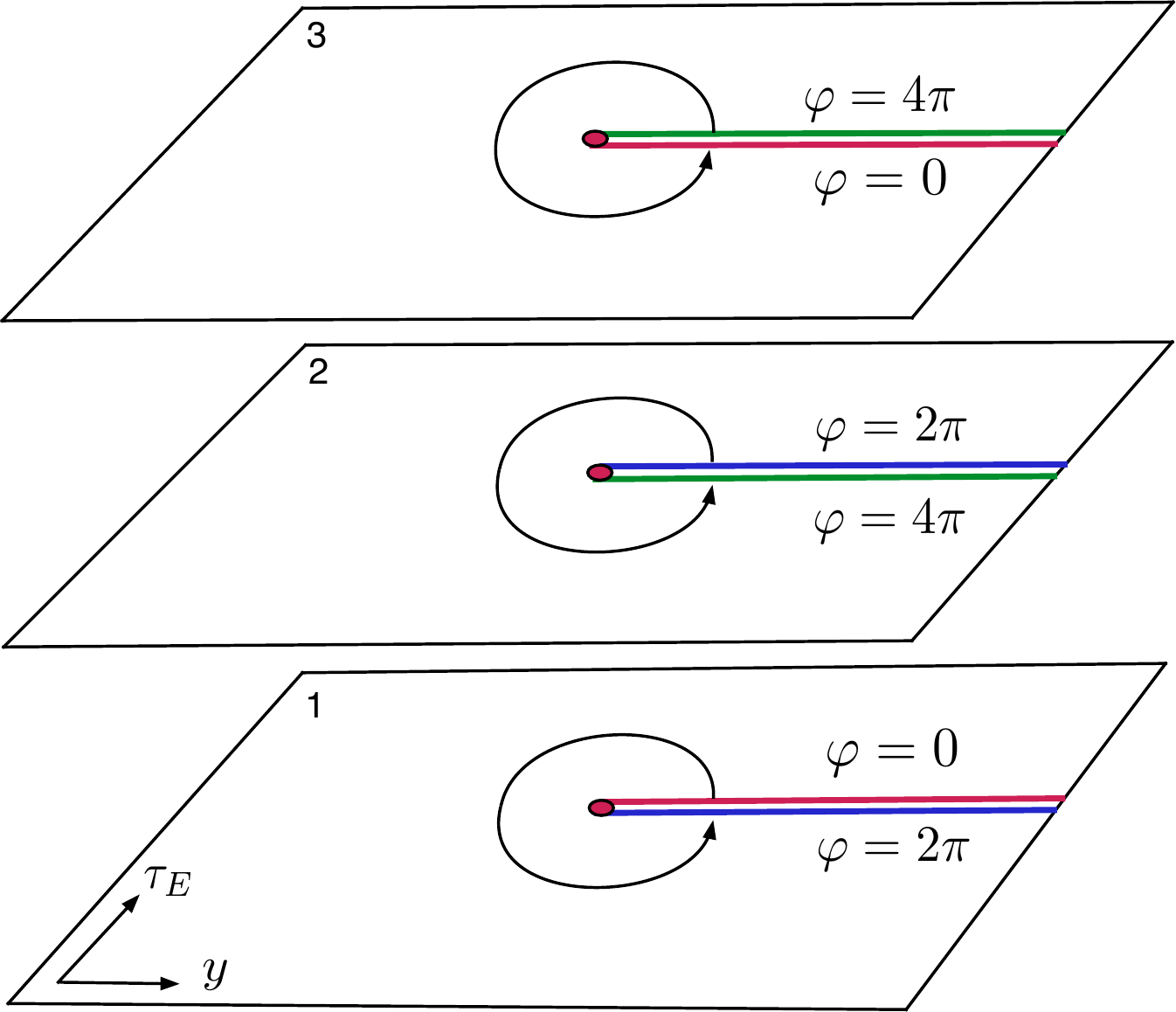}{9}{\small{\textsf{The replica manifold for $n=3$ has three sheets (the $x$ direction is implicit) which are glued along $A$ as indicated -- the blue line on sheet one is glued with the blue line on sheet 2 and so on. In this case $\varphi \in [0,6\pi)$ with $\varphi=6\pi$ identified with $\varphi=0$. As $\varphi$ goes from $2\pi (k-1)$ to $2\pi k$, we cover the $k$th sheet.}}}

Then, we have
\beq
S_{n,bulk} = \frac{1}{1-n}\Big(\mathrm{ln}\;Z_{bulk}(M_n)-n\;\mathrm{ln}\;Z_{bulk}(M_1)\Big)
\eeq
where $Z_{bulk}(M_n)$ is the partition function of the bulk theory on the replica manifold $M_n$, and $M_1$ is the original spacetime $\re^3$. 

Now we want to ask the same question -- what happens to the $n$th R\'{e}nyi entropy under the local diffeomorphism \eqref{Ediff} around the entangling surface? From the above expression, we find that the Renyi entropy changes as
\beq\label{Renyivar}
\delta_{\xi} S_{n,bulk} = \frac{1}{1-n}\Big(\delta_{\xi}\mathrm{ln}\;Z_{bulk}(M_n) -n\delta_{\xi}\mathrm{ln}\;Z(M_1)\Big).
\eeq
We now use the fact that in parity-violating theories, the effective action has a gravitational Chern-Simons term
\beq
-\mathrm{ln}\;Z_{bulk}(M_n)= \int_{M_n}\frac{i\kappa(x)}{2}\;\left({\Gamma^{\alpha}}_{\beta}\wedge d{\Gamma^{\beta}}_{\alpha}+\frac{2}{3}{\Gamma^{\alpha}}_{\beta}\wedge{\Gamma^{\beta}}_{\gamma}\wedge {\Gamma^{\gamma}}_{\alpha}\right)+\cdots
\eeq  
where the ellipsis indicates other terms which might be present, but do not contribute to the present computation.\footnote{For instance, one might worry about the parity-odd Hall viscosity term \cite{Hughes:2012vg} which was mentioned in footnote \ref{HVfootnote}. We now see that this term, though present, cannot contribute to the present computation because it is diffeomorphism invariant.}  Under a coordinate transformation, the Christoffel connection $\Gamma$ transforms as
\beq
\delta_{\xi}{\Gamma^{\alpha}}_{\mu\beta} = \nabla_{\mu}\left(\pa_{\beta}\xi^{\alpha}\right)
\eeq
and consequently the Chern-Simons action transforms as
\beqn
\delta_{\xi}\mathrm{ln}\;Z_{bulk}(M_n) &=& \frac{i}{2}\int_{M_n}d\kappa \wedge \left(\pa_{\beta}\xi^{\alpha}\right)d{\Gamma^{\beta}}_{\alpha}\label{Renyi1}\\
&=&-\frac{i}{2}\int_{M_n}d\kappa \wedge \xi^{\alpha}d\left(\pa_{\beta}{\Gamma^{\beta}}_{\alpha}\right)-\frac{i}{2}\int_{M_n}d\left(\pa_{\beta}\kappa\right) \wedge \xi^{\alpha}d{\Gamma^{\beta}}_{\alpha}\label{Renyi2}
\eeqn
where we have consistently dropped all boundary terms, because $\xi\to 0$ at infinity. The integration by parts in going from \eqref{Renyi1} to \eqref{Renyi2} is meaningful as long as the geometry of the replica manifold is appropriately regulated (see below). The first term in \eqref{Renyi2}, in comparison with equation \eqref{consanomaly}, makes it clear that the entanglement response to boosts is directly related to the \emph{consistent} gravitational anomaly of the domain wall CFT. The second term in \eqref{Renyi2} vanishes simply because the index $\beta$ has to be along the domain wall for the Christoffel symbol to be non-trivial, but then $\pa_{\beta}\kappa=0$. Hence, we conclude that
\beq
\delta_{\xi}\mathrm{ln}\;Z_{n, bulk}=-\delta_{\xi}\mathrm{ln}\;Z_{n, CFT}
\eeq
which directly implies that the anomalous response of the bulk entanglement entropy cancels the anomalous response of the boundary. Since the anomaly above is the consistent anomaly, the gravitational Chern-Simons term -- which is not diffeomorphism invariant -- is the only relevant term in the bulk (see \cite{AlvarezGaume:1984dr,AlvarezGaume:1984nf,Hughes:2012vg} for a discussion of consistent \emph{v}. covariant anomalies, and their relation to Chern-Simons terms).

Now all that remains is to substitute the above expression into equation \eqref{Renyivar}. Let us introduce polar coordinates in the $(\tau_E, y)$ plane  
\beq
r = \sqrt{\tau_E^2+y^2},\;\;\varphi = \mathrm{tan}^{-1}\frac{\tau_E}{y}.
\eeq
With a slight abuse of notation, we take $\varphi\in [0,2\pi n)$ -- when $\varphi$ crosses $2\pi k$ (for $k\in \mathbb{Z}$, $1\leq k \leq n-1$) we move from the $k$-th replica sheet to the $(k+1)$-th sheet. Now, the full $n$-sheeted replica manifold $M_n$ can be regarded as the space $\re^{3}$ with the metric
\beq
g_n = dr^2+n^2r^2d\theta^2+dx^2 \label{conemetric}
\eeq
where we have defined $\varphi = n\theta$, with $\theta \in [0,2\pi)$. This metric has a conical singularity along the line $r=0$ which needs to be regulated. We use the regularized metric \cite{Solodukhin:1994yz} (see Fig. 6 below)
\beq
g_{n,a}=\frac{r^2+a^2n^2}{r^2+a^2}dr^2+n^2r^2d\theta^2+dx^2.
\eeq
Clearly, in the limit $a\to 0$ we recover the original cone metric \eqref{conemetric}. 
\begin{figure}[!h]
\centering
\begin{tabular}{l l l l}
\includegraphics[height=3.2cm]{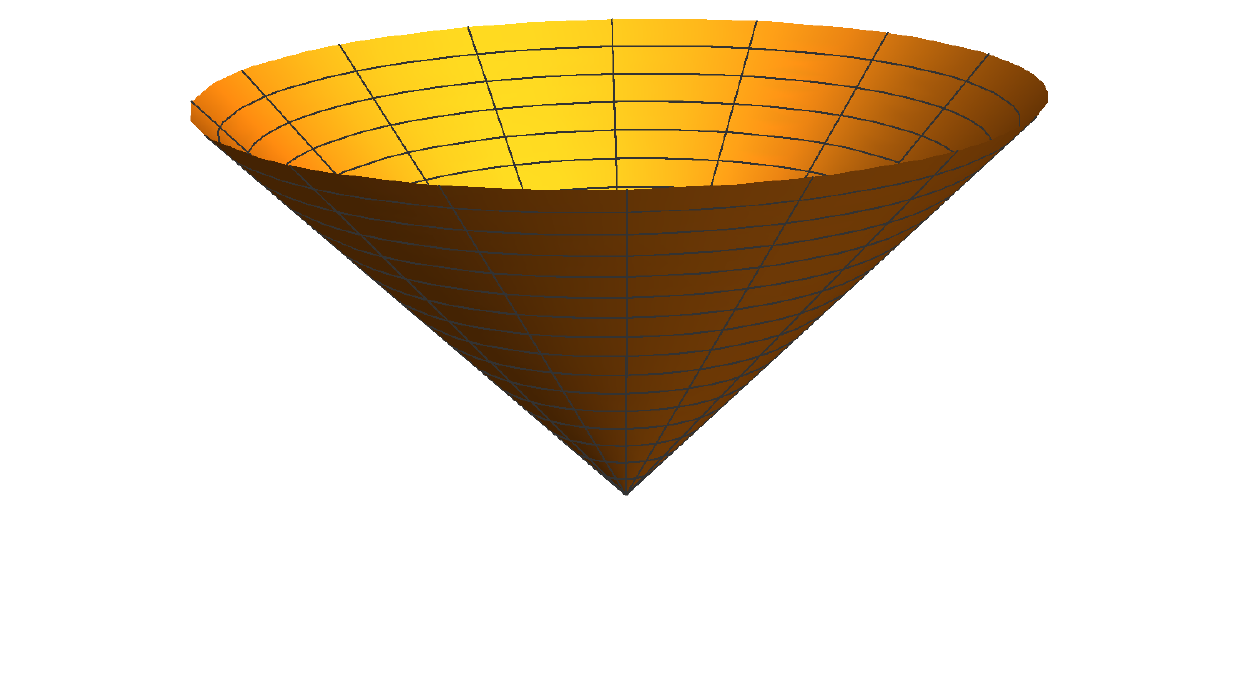} & \includegraphics[height=3.1cm]{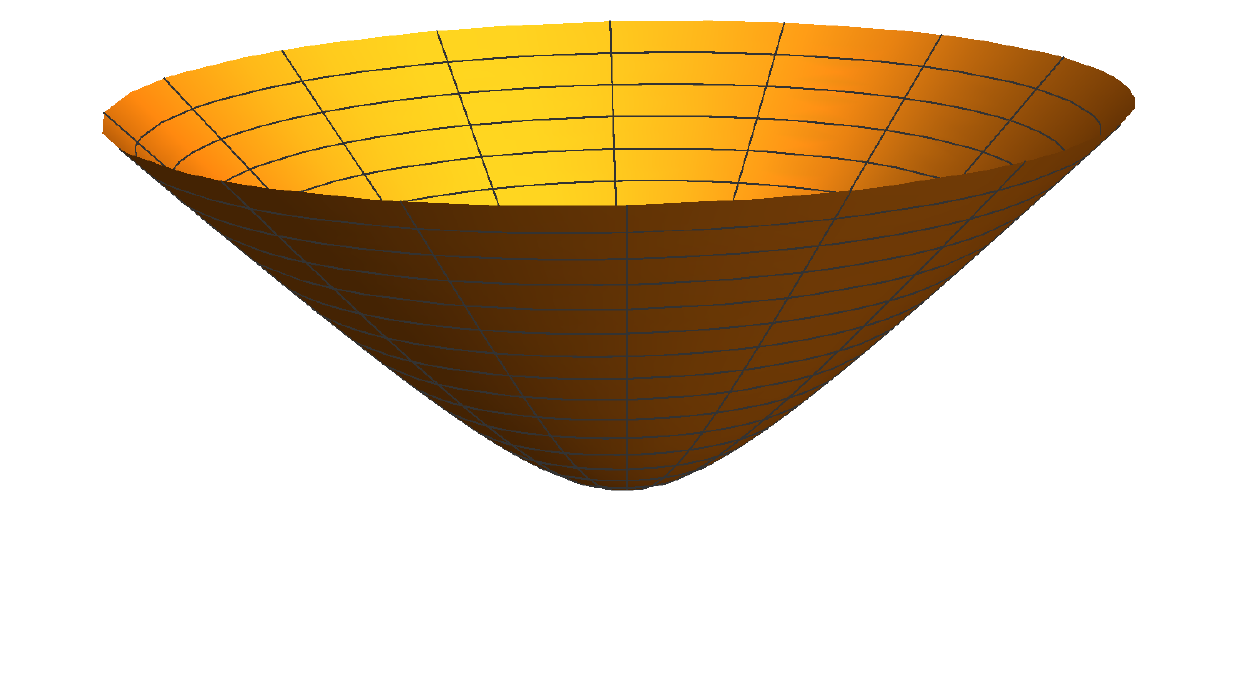}
\end{tabular}
\caption{\small{\textsf{(Left) The geometry of the replica manifold in the $(\tau_E,y)$ directions is that of a cone embedded in $\re^{1,2}$. (Right) We can regularize the conical singularity at origin in the $(\tau_E,y)$ plane by replacing it with a smoothed-out cone.}}}
\end{figure}
We find that the non-trivial Christoffel symbols for the regularized metric are given by
\beq
{\Gamma^r}_{\theta\theta} = -\frac{r(r^2+a^2)n^2}{r^2+n^2a^2},\;\;{\Gamma^r}_{rr} = -\frac{a^2r(n^2-1)}{(r^2+a^2)(r^2+n^2a^2)},\;\;{\Gamma^{\theta}}_{r\theta} = \frac{1}{r}.
\eeq
After substituting these into equation \eqref{Renyi2}, and using $\xi = \frac{\eta_E\phi(\Bx)}{n} \pa_{\theta},$ one finds from a straightforward calculation
\beq
\delta_{\xi}\mathrm{ln}\;Z_{bulk}(M_n) = 
i\pi\eta_E (\kappa_+-\kappa_-)\frac{(1-n^2)}{n}.
\eeq
Putting this into equation \eqref{Renyivar} and using \eqref{BBC}, we find\footnote{The Renyi entropies in topological field theories are typically $n$-independent. Here, the $n$-dependence for the bulk entanglement response clearly comes from the presence of the domain wall defect. Indeed, it precisely cancels the contribution coming from the domain wall CFT.}
\beq
\delta_{\xi}\;S_{n,bulk} = i\eta_E \frac{c_L-c_R}{48}\left(1+\frac{1}{n}\right).
\eeq
Finally, taking the limit $n\to 1$ and reverting to Lorentzian signature, we obtain
\beq
\delta_{\xi}S_{EE,bulk} =  \eta \frac{(c_L-c_R)}{24},
\eeq
which indeed matches our calculation in the previous section. Additionally, we repeated the above calculation using Cartesian coordinates (instead of polar coordinates) for the replica manifold, and we found precisely the same result stated above.\footnote{It has been pointed out in \cite{IqbalWall} that it might be important to treat the coordinate singularities at the entangling surface as boundaries, possibly leading to extra contributions.} In a way, this calculation makes transparent the fact that the cancellation of the anomalous entanglement response of the bulk and the domain wall CFT is forced upon us by the diffeomorphism invariance of the full theory on replica manifolds.

\subsection{Entanglement inflow}\label{sec2.3}
The calculations in the previous section, while concrete, perhaps hide the physics behind the matching of the bulk and domain wall CFT results. We will now describe the bulk-boundary cancellation of the anomalous entanglement entropy using a more intuitive, albeit less rigorous picture of flow of entanglement entropy (or equivalently entanglement `energy') which emerges in the Rindler wedge. The \emph{Rindler wedge} is the domain of dependence of the region $A$ (where recall that $A$ in our calculation is the half space $y>0$ on the spatial slice at $t=0$) over which the reduced density matrix is defined (see Fig.~\ref{fig:fig7.pdf}a).

\myfig{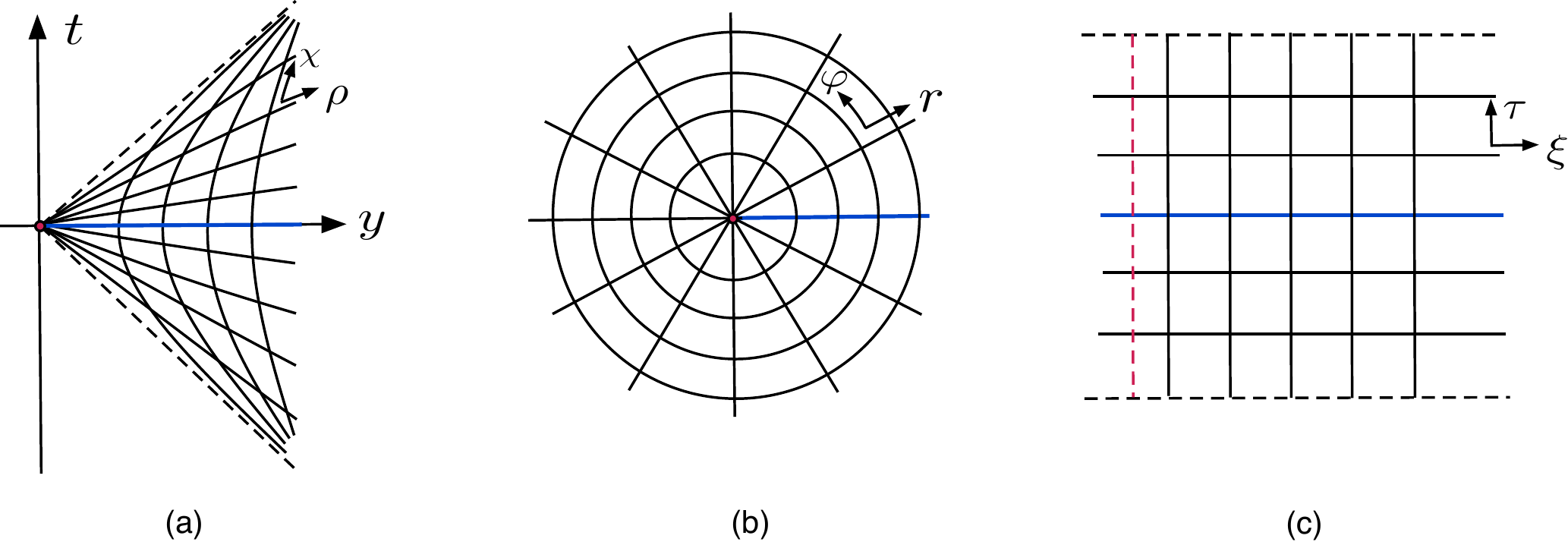}{14}{\small{\textsf{(a) A constant $x$ slice of the Rindler wedge. The modular Hamiltonian generates translations along $\chi$. (b) In Euclidean signature, the modular Hamiltonian generates rotations around the entangling surface, i.e., translations in $\varphi$. (c) A constant $x$ slice of $S^1\times \mathbb{H}^2$, with $\xi = \mathrm{ln}(z)$. The $\tau$ direction is compact. The red dashed line represents the cutoff at $z=a$. The bold blue line represents the region $A$ over which the reduced density matrix is defined.}}}

Therefore, all of the physics relevant to the region $A$ can be equivalently described by going to Rindler coordinates
\beq
\chi = \mathrm{tanh}^{-1}\left(\frac{t}{y}\right),\;\;\rho = \sqrt{y^2-t^2},
\eeq
where the transverse coordinate $x$ remains intact. The Minkowski metric in these coordinates is given by
\beq
g_{Rindler} = -\rho^2d\chi^2+d\rho^2 +dx^2.
\eeq
Importantly, the \emph{modular/entanglement Hamiltonian} in this picture is proportional to the generator of boosts, or $\chi$ translations. It is more convenient to consider the Euclidean version of this statement. Consider Wick-rotating from Minkowski space $\re^{1,2}$ to Euclidean space $\re^3$ with the metric 
\beq
g_{Eucl.} = r^2d\varphi^2+dr^2 +dx^2.
\eeq
with $\varphi \in [0,2\pi)$ being the polar angle in the $(r,\varphi)$ plane (See Fig.~\ref{fig:fig7.pdf}b). In this case, the entanglement Hamiltonian is exactly $2\pi$ times the generator of rotations in the $(r,\varphi)$ plane, or in other words $\varphi$-translations (see Appendix \ref{appA} for a derivation). It is further convenient to transition from $\re^3$ to the conformally equivalent space $S^1\times \mathbb{H}^2$ (where $\mathbb{H}^2$ stands for two-dimensional hyperbolic space). Let us use the coordinates\footnote{The coordinate $\tau$ should not be confused with the Euclidean time coordinate $\tau_E$.} $(\tau, z,x')$ to parametrize $S^1\times \mathbb{H}^2$, with $\tau \in [0,2\pi),\;z\in (0,\infty)$ and $x'\in \re$ (Fig.~\ref{fig:fig7.pdf}c). In these coordinates, the metric on $S^1\times \mathbb{H}^2$ is given by
\beq
g_{S^1\times \mathbb{H}^2} = d\tau^2+\frac{dz^2+dx^2}{z^2}.
\eeq
The map $\re^3 \mapsto S^1\times \mathbb{H}^2$ in terms of the coordinates $(\varphi,r, x)$ on $\re^3$ and the coordinates $(\tau,z,x')$ on $S^1\times\mathbb{H}^2$ is simply
\beq
\tau(r,\varphi,x) = \varphi,\;\;z(r,\varphi,x) = r,\;\;x'(r,\varphi,x)=x.\label{CT}
\eeq 
Since the $x$ coordinate maps trivially to $x'$, we will drop the prime from now on for simplicity. It is clear that \eqref{CT} is a conformal transformation. This fact has been used quite successfully in studying the entanglement entropy of conformal field theories \cite{Casini:2011kv, Faulkner:2014jva}. In our case, the domain wall theory is indeed a conformal field theory (in the limit of infinitesimal wall thickness), but in the bulk we do not have a CFT. However, the object of primary focus in the bulk is the parity-violating, gravitational Chern-Simons term in the bulk effective action -- this term is indeed conformally invariant (up to boundary terms). Further, in many condensed matter systems of interest, such as topological phases of matter, (e.g., topological insulators\cite{haldane1988,hasan2010} and fractional quantum Hall systems\cite{wenreview}, etc.) the bulk is effectively described by a topological field theory at low energies, which is certainly conformally invariant, while the corresponding edge theories, at first approximation, are gapless CFTs. Hence, we expect the description in this section to apply to a wide class of such topological systems. 

The reason it is convenient to transition to $S^1\times \mathbb{H}^2$ is as follows: the region $A$ over which the reduced density matrix is defined maps to a constant-$\tau$ slice in $S^1\times \mathbb{H}^2$. The modular/entanglement Hamiltonian $\widehat{H}_E$ now generates `time translations' in the $\tau$ direction, and thus can be regarded as $2\pi$ times the \emph{physical} Hamiltonian on hyperbolic space:
\beq
\widehat{H}_E= 2\pi\int_{-\infty}^{\infty}\frac{dx}{z}\int_{0}^{\infty}\frac{dz}{z}\;\hT_{\tau\tau}(0,z,x)+\mathrm{constant}.
\eeq
Therefore, the reduced density matrix $\rho_A = e^{-\widehat{H}_E}$ is precisely the \emph{thermal} density matrix with inverse temperature $\beta = 2\pi$, and so in hyperbolic space-time, the entanglement entropy is  nothing but the thermal entropy. Hence, the first law of entanglement entropy\cite{Blanco:2013joa,Faulkner:2013ica} 
\beq
\delta S_{EE} = \delta E \equiv \delta \left\langle \widehat{H}_E\right\rangle=\delta\; \mathrm{Tr}_{\mathcal{H}_{A}}\left(\widehat{H}_Ee^{-2\pi\widehat{H}_E}\right).
\eeq
naturally takes the form of the first law of thermodynamics. Therefore, from this point of view the anomalous change in the entanglement entropy due to a local boost is equivalent to a change in the total energy of the system. This is quite remarkable since it is an explicit map between the variation of entanglement entropy and variation of ``energy." Now, due to the gravitational anomaly of the domain wall CFT, its intrinsic energy is not conserved, which (by the first law) is reflected in the anomalous variation of its intrinsic entanglement entropy. However, in the full bulk plus domain wall theory, the total energy must be conserved, and as such, the bulk must somehow account for the energy lost by the domain wall. We claim that this happens as follows: in addition to the chiral states on the domain wall, there are also chiral states localized on the (regularized) bulk entangling surface. Being chiral, their energy is also not conserved during a Lorentz boost, but in precisely the right way to account for the energy non-conservation of the domain wall. Hence there must be anomaly matching between the physical domain wall/edge and the regularized entangling surface.

To see this in more detail, let us first focus on the CFT living on the domain wall. The domain wall corresponds to the slice of $S^1\times \mathbb{H}^2$ at $x=0$, which is actually a cylinder with the metric
\beq
g_{DW}= d\tau^2+\frac{dz^2}{z^2}=d\tau^2+d\xi^2,
\eeq
where in the second equality we have defined $\xi = \mathrm{ln}(z)$. The local boost of the Cauchy surface now corresponds to a local translation in the $\tau$ direction (see Fig.~\ref{fig:fig8.pdf})
\beq
\xi = \eta_E\phi(z,x)\;\pa_{\tau}
\eeq
where recall that $\phi$ is a bump function supported over the ball $B_R$ around $z=x=0$ (with $R$ much larger than the thickness of the domain wall, or any short distance cut-offs in the theory). Ordinarily, this boost would conserve energy (because the Hamiltonian is a conserved charge) provided there is no flux through the boundary of hyperbolic space at $z=0$, which is really infinitely far away. However, in practice we must regulate this infinity by putting a cutoff at some finite value $z=a$.\footnote{As with all aspects of entanglement calculations, it is essential to regulate the theory. The cutoff at $z=a$ maps back to the exclusion of a small cylindrical region around the entangling surface, which is a UV cutoff.} Then, owing to the chirality of the CFT, the flux condition is clearly violated because of the energy-flux through the window between $\tau=0$ and $\tau=\eta$ at $z=a$. Of course, if the theory is non-chiral such that $c_L=c_R$ then the outgoing flux cancels the incoming flux, and the energy remains unchanged.

\myfig{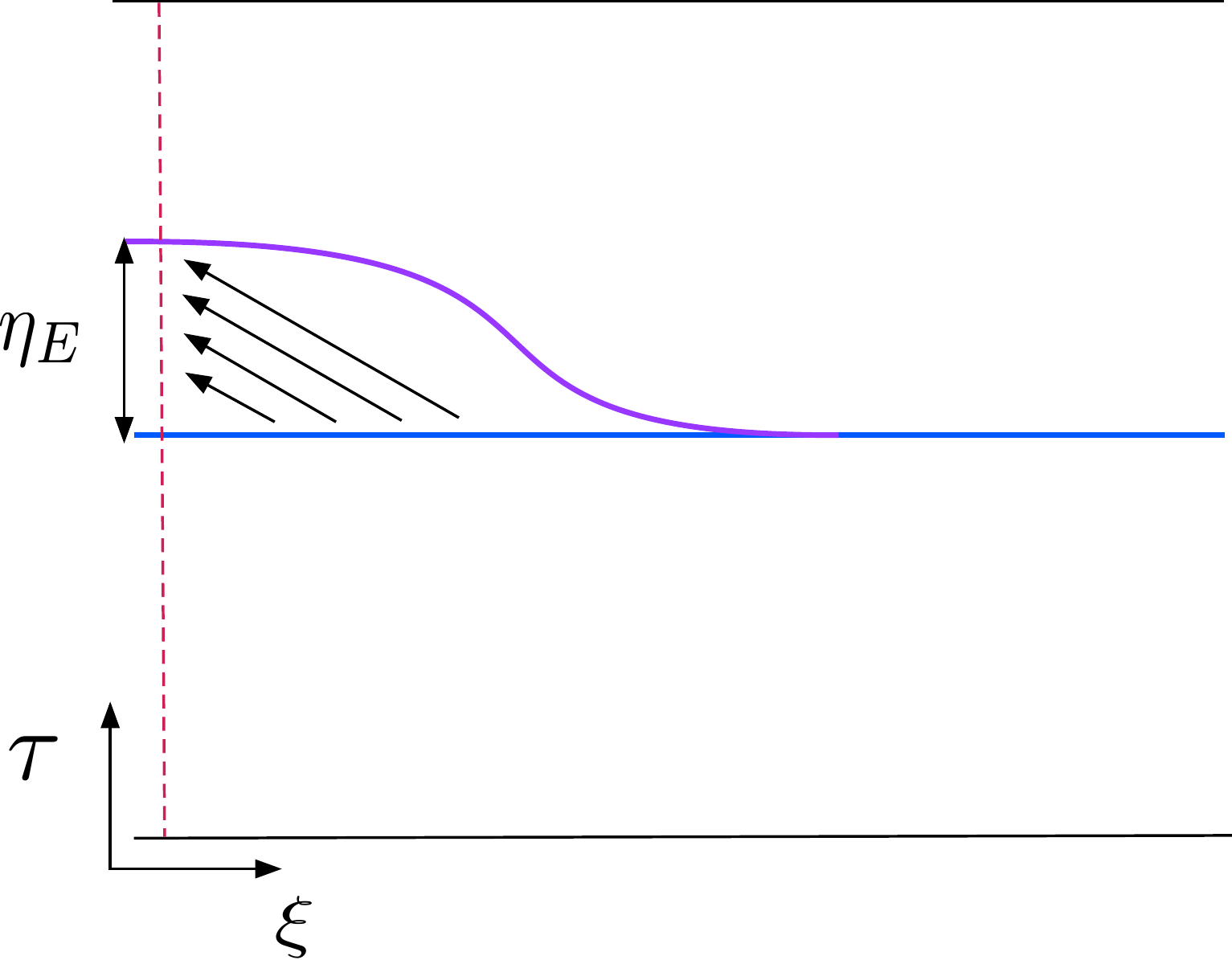}{7}{\small{\textsf{The local boost corresponds to a local $\tau$-translation in $S^1\times \mathbb{H}^2$. The blue line is the original constant $\tau$ surface over which we measure the energy, while the purple line is the new surface after the boost. The arrows indicate that in a chiral CFT, we can have an energy-flux through the cutoff surface (dashed red line) during such a deformation.}}}

It is straightforward to quantify how much energy is lost/gained during this deformation. We know that the expectation values of the holomorphic and anti-holomorphic stress tensors on a cylinder are given by
\beq
2\pi\langle \hT\rangle_{cylinder} = -\frac{c_L}{24},\;\;2\pi \langle \widehat{\bar{T}}\rangle_{cylinder} = -\frac{c_R}{24}
\eeq
We can then use the above results to compute the flux through infinity, and one finds 
\beq\label{fluxDW}
2\pi \delta E_{DW} = 2\pi\int_0^{\eta_E}d\tau\;\langle \hT_{\tau \xi}(\tau,\mathrm{ln}(a))\rangle =  2\pi i \int_0^{\eta_E}d\tau\left(\langle \hT\rangle-\langle \widehat{\bar{T}}\rangle\right)=-i\eta_E\frac{c_L-c_R}{24}
\eeq
where the subscript $DW$ refers to domain wall. We see that the above result is essentially proportional to the \emph{Casimir momentum} density along the domain wall. The first law of entanglement entropy then gives
\beq
\label{eq:eedge}
\delta S_{EE,DW} = -i\eta_E\frac{c_L-c_R}{24}=-\eta\frac{c_L-c_R}{24}
\eeq
So we conclude that the change in the domain wall entanglement entropy is caused by energy flowing out from the boundary at $z=a$. 

Let us now revisit the situation from the bulk point of view (see Fig.~\ref{fig:fig9.pdf}). As we discussed above, since UV divergences arise from the infinite amount of space close to the bulk entangling surface $z=0$, it is necessary to regulate the geometry in this region. Here we choose to put a hard cutoff at $z=a$ (which can be thought of as a stretched Rindler horizon), analogous to the brick-wall model of black hole thermodynamics \cite{'tHooft1985727}; another possibility is to use a lattice regularization. In such a regularization, it is expected that there must also exist chiral states localized on the cutoff surface $z=a$, in addition to the chiral states on the domain wall $x=0$ \cite{lihaldane,PhysRevLett.101.010504,fidkowski2010,turner2010,sterdyniak2011,PhysRevB.84.195103,chandran2011,hughes2011,PhysRevLett.108.196402,PhysRevB.86.045117}. The existence of these states is forced upon us by the requirement of anomaly cancellation for the theory on $S^1\times \mathbb{H}^2$ (or equivalently on the Rindler wedge). In particular, the modes supported in the region $x<\frac{\ell}{2}$ of the entangling surface must have net chirality $c_L-c_R=48\pi\kappa_-$, while those in the region $x>\frac{\ell}{2}$ must have chirality $c_L-c_R=48\pi\kappa_+$, where $\ell$ is the thickness of the domain wall. It is possible to demonstrate the existence of these states explicitly for free fermions by solving the Dirac equation on $\re\times \mathbb{H}^2$ with brick-wall boundary conditions; this is done in Appendix \ref{appES}. However, we reiterate that they are guaranteed by diffeomorphism invariance (i.e., anomaly cancellation). These states present a natural resolution to the apparent loss of energy from the domain wall -- \emph{the energy which is lost in the domain wall CFT at $z=a$ is simply transferred to the chiral states localized on the regularized entangling surface in the bulk}.

\myfig{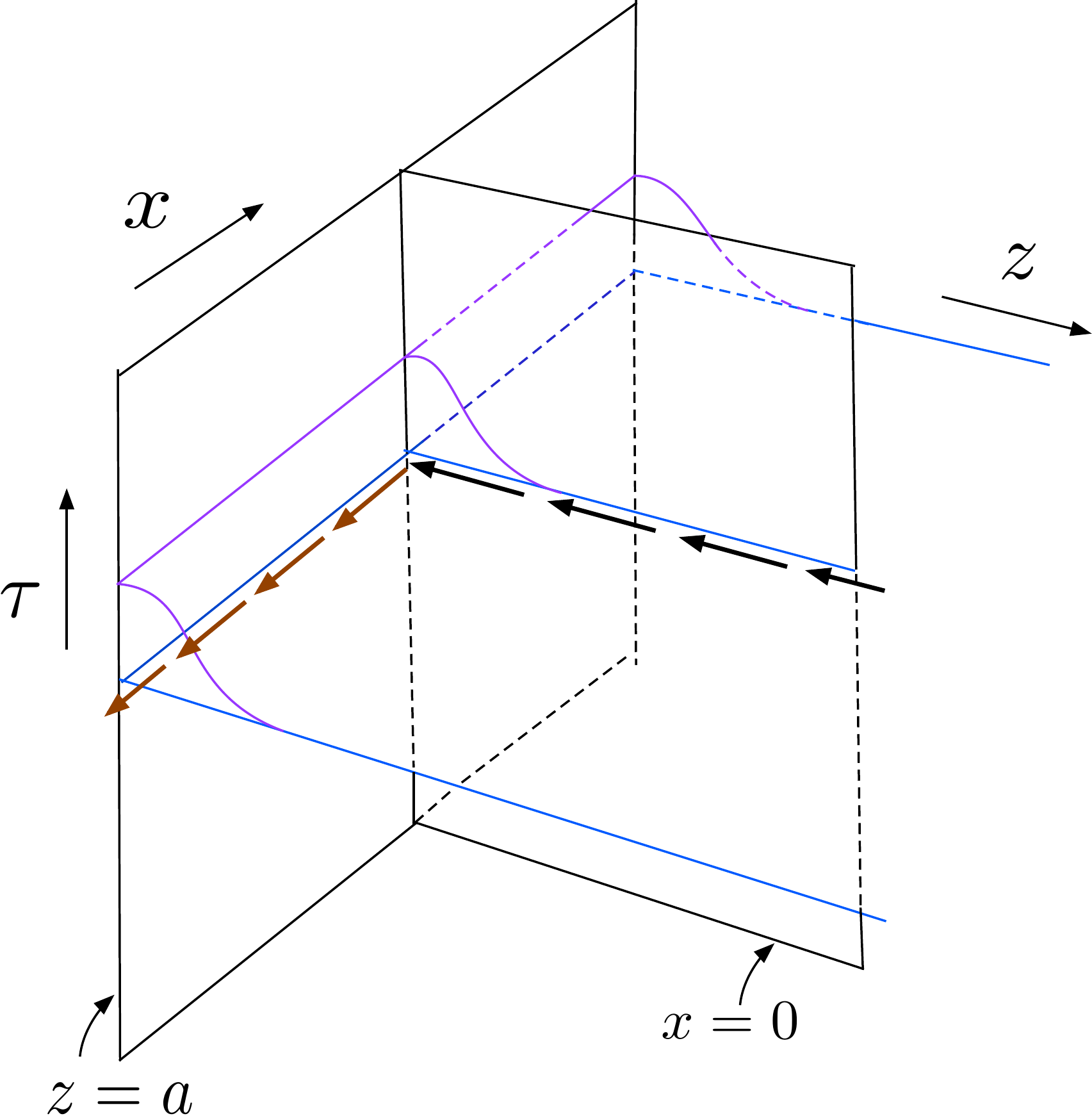}{8}{\small{\textsf{The boost from the bulk point of view for the case $\kappa_+=0$, $\kappa_- \neq 0$: the blue plane is the original constant $\tau$ surface over which we measure the energy, while the purple surface is the new surface after the boost. The black and brown arrows indicate the energy-flux associated with edge states localized on the domain wall and bulk entangling surface respectively. The energy flux which leaves the domain wall enters the bulk along the entangling surface. The direction the edge states flow along the entangling surface depends not only on the nature of the domain wall (i.e. $(\kappa_+-\kappa_-)$) but on the way the bulk  theories are regularized on opposite sides of the domain wall (i.e. the specific values $\kappa_+$ and $\kappa_-$). For this figure we have chosen a regularization where $\kappa_{+}=0$ and $\kappa_{-}\neq 0.$}}}

In order to confirm this, we may approximately describe the chiral edge states localized on the regularized entangling surface as a CFT on the cutoff surface $z=a$. We now compute the energy flux through a constant $x$ slice 
\beq
\int_0^{\eta_E}d\tau\;\left\langle \hT_{\tau x}(\tau,x)\right\rangle
\eeq
along the cylinder at $z=a$, for $x < -\ell/2$ and $x>\ell/2$.\footnote{In the limit $a\to 0$, the metric on the cutoff surface $z=a$ is conformally equivalent to the degenerate metric $\mathrm{lim}_{z\to 0}\left(dx^2+z^2d\tau^2\right)$. However, the computation of the flux of Casimir momentum is insensitive to this degeneracy.} Repeating the arguments given above involving the Casimir momentum density, we find that the change in the energy of the region $x<-\frac{\ell}{2}$ of the cutoff surface is given by
\beq
\delta E_{ES}(x<-\frac{\ell}{2})=-2\pi i\eta_E\kappa_-
\eeq
while for the region $x>\frac{\ell}{2}$, we have
\beq
\delta E_{ES}(x>\frac{\ell}{2})=+2\pi i\eta_E\kappa_+
\eeq
where we have used the subscript $ES$ to indicate that these quantities are computed along the cutoff entangling surface. Therefore, the total change in the energy during the boost, coming from the entangling surface is
\beq
\delta E_{ES} =  2\pi i\eta_E(\kappa_+-\kappa_-) = \eta\frac{c_L-c_R}{24}
\eeq
which precisely accounts for the energy-flux out of the domain wall \eqref{fluxDW}. By the first law of entanglement entropy, we thus conclude that $\delta S_{EE, ES} = -\delta S_{EE, DW}$. Note that this result does not depend on much of the detailed properties of the spectrum of modes on the entangling surface.

The chiral states which mediate the inflow of energy flux between the bulk and domain wall essentially constitute the low-energy spectrum of the entanglement Hamiltonian $\widehat{H}_E$. We have noted above that since this is \emph{the} Hamiltonian in the above Rindler space construction, the diffeomorphism invariance of the full theory forces the existence of these chiral states in its spectrum. Indeed, in materials exhibiting topological phases of matter, it is known that there is a correspondence between the low-lying spectrum of the entanglement Hamiltonian and the low-energy spectrum of the physical Hamiltonian in the presence of a physical boundary\cite{lihaldane,PhysRevLett.101.010504,fidkowski2010,turner2010,sterdyniak2011,PhysRevB.84.195103,chandran2011,hughes2011,PhysRevLett.108.196402,PhysRevB.86.045117}. We have seen above that the cancellation of the bulk and boundary responses to a local boost can be explained as an inflow of (entanglement) energy associated exactly with these low-lying states in the entanglement spectrum, which we interpret as an \emph{inflow of entanglement entropy}.

\myfig{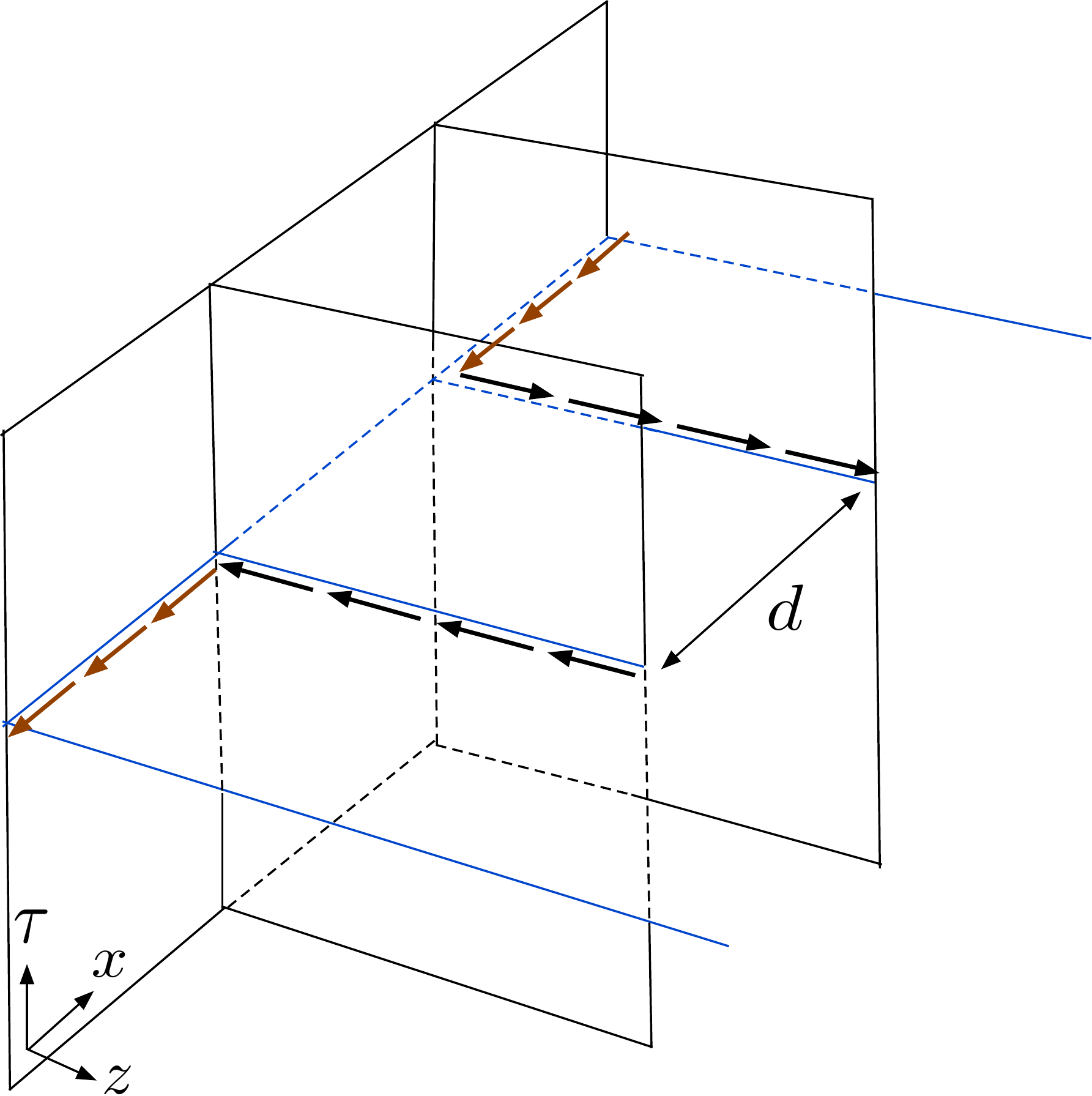}{8}{\small{\textsf{Two domain walls with opposite chiralities at a distance $d$. The arrows indicate energy-flux along the domain walls and the bulk entangling surface. In the limit $d\to 0$, the walls can be gapped out, but we are still left with chiral states localized on the cutoff entangling surface.}}}

Interestingly, we could also reverse this logic to deduce that the conservation of energy in the Rindler wedge forces upon us the existence of chiral edge states localized on the entangling surface to absorb the anomaly. In fact, we can now see that this conclusion is actually independent of the domain wall construction, and is true whenever a spatial entanglement cut passes through a topological phase, i.e., when the entanglement cut passes through a gapped region with a topological effective action. For example,  consider the situation in Fig.~\ref{fig:fig10.pdf} where we have two domain walls of opposite chiralities. Now let us bring the two domain walls closer such that they eventually annihilate --- i.e.,  the limit where the two walls overlap and the low-energy modes can tunnel between each wall. In this limit there is no net chirality on the wall, and the domain wall states can be locally gapped. Interestingly,  even in this situation, we are still left with chiral states along the entire entangling surface away from the wall, as should be clear from Fig.~\ref{fig:fig10.pdf}. As the wall region is shrunk to zero this shows that the entire entanglement cut will bind low-lying chiral entanglement modes. This gives us an intuitive way to understand why the low-lying spectrum of the entanglement Hamiltonian in a topological phase is in close correspondence to the edge spectrum in the presence of a physical boundary; at the very least the low-lying spectrum of the modular Hamiltonian must be able to absorb the anomalous response of the domain wall/edge states.

\subsection{Explicit calculation for a Chern insulator lattice model}\label{sec2.4}

We now present some explicit lattice model calculations to demonstrate the above physics in the context of a condensed matter system called the Chern insulator \cite{haldane1988}. The Chern insulator is a 2+1-d lattice model of massive Dirac fermions which can generate a nonzero bulk Chern number in certain parameter regimes. A non-vanishing Chern number $C_1$ indicates that the system will have a bulk quantum Hall effect with Hall conductance $\sigma_{xy}=\tfrac{e^2}{h}C_1,$ and thus will have chiral edge states with $c_{L}-c_{R}=C_1.$ 

The 2+1-dimensional lattice model we consider has discrete translational symmetry with two degrees of freedom per site. The real-space model can be Fourier transformed to generate the Hamiltonian
\begin{equation}
H=\sum_{k}c^{\dagger}_{a k}H_{ab}(k)c_{b k},
\end{equation}\noindent where $a,b=1,2.$ The momentum-dependent matrix $H_{ab}(k)$ is the Bloch Hamiltonian. 
To be concrete, we consider the following single particle Bloch Hamiltonian: 
\begin{equation}
\label{eq:CI}
H(k)=v_x\sin k_{x}\sigma^{x}+v_y\sin k_{y}\sigma^{y}+(2-m-\cos k_{x}-\cos k_{y})\sigma^{z}
\end{equation}\noindent where we have set the lattice constants $a_{x},a_{y}=1,$ $\hbar=1,$ $k_{x},k_{y}$ label the lattice momenta, $m, v_x, v_y$ are parameters, and we assume that $v_x, v_y >0.$ The two bulk bands of this model have energies $E_{\pm}=\pm\sqrt{v_{x}^2\sin^2 k_x+v_{y}^2 \sin^2 k_y +(2-m-\cos k_{x}-\cos k_{y})^2}$. 
This Hamiltonian is a gapped topological insulator with $C_1=-1$ in the regime $0< m< 2,$ or with $C_1=+1$ in the regime $2< m< 4.$ Since a system with a nonzero Chern number exhibits localized chiral edge states, it can be used to test our hypothesis that the parity-odd change in the entanglement entropy of a chiral fermion is compensated by the change in the bulk entropy when boosted with a boost parameter $\eta$.

Let us now try to see how we can manipulate the parameters in this system to get something akin to a boost as in the relativistic cases we have been discussing in previous sections. Consider the system in a spatial cylinder geometry with open boundaries in $x$, and periodic boundaries in $y$ (see Fig. \ref{fig:fig11.pdf}). The dispersion of the boundary fermions localized on the $x$ edge for this model is given by $v_y\sin k_{y}$\cite{creutz2001}. We can thus infer that the number $v_y$ determines the velocity of the low-energy chiral fermions. 
Indeed, the low-energy Hamiltonian of a chiral fermion localized on the $x$-edge of the system in eq.~\eqref{eq:CI} is given by $H=v_y k_{y}$, when expanded near $k_y=0.$ Under an infinitesimal boost as in eq.~\eqref{eq:boostdiff}, and very close to the origin in Fig.~\ref{fig:fig3.pdf}, we can calculate the change in the Hamiltonian by calculating $\delta_{\xi}H=\left\lbrack H,\xi\right\rbrack$ with $\xi$ given by $-i\eta(yp_{t}+tp_{y})$. 
\begin{eqnarray}
\label{eq:boost}
\delta_{\xi}H&=&-i\left(\frac{\partial H}{\partial x^{\alpha}}\frac{\partial \xi}{\partial p_{\alpha}}-\frac{\partial H}{\partial p_{\alpha}}\frac{\partial \xi}{\partial x^{\alpha}}\right)\\\nonumber
&=&i\frac{\partial H}{\partial p_{t}}\frac{\partial \xi}{\partial t} 
\end{eqnarray}\noindent where $\alpha=t,x,y$. Using the fact that $\partial H/\partial p_{t}=v_{y}$, we see that the new Hamiltonian in the boosted frame at first order in $\eta$ is given by $H_{boost}=v_y(1+\eta)p_{y}$. Thus, a boost acts just like  a change in the velocity of the low energy limit of the chiral fermions, and we will interchangeably use $\delta v/v_{y}$ and $\eta$ from now on in this section. 

Now let us consider the entanglement of this system. Since the bulk model is gapped (except at the edges) and does not have topological ground state degeneracy, it is expected to have an entanglement entropy which scales linearly with the length of the entanglement cut (passing through a gapped region), without a universal sub-leading correction:  
\begin{equation}
\label{eq:eeunchanged}
S_{CI}=\alpha L,
\end{equation}\noindent where $L$ is the length of the entanglement cut and $\alpha$ is a constant which will generally depend on the parameters in the model, and on the boundary conditions that are imposed on the system. 

\myfig{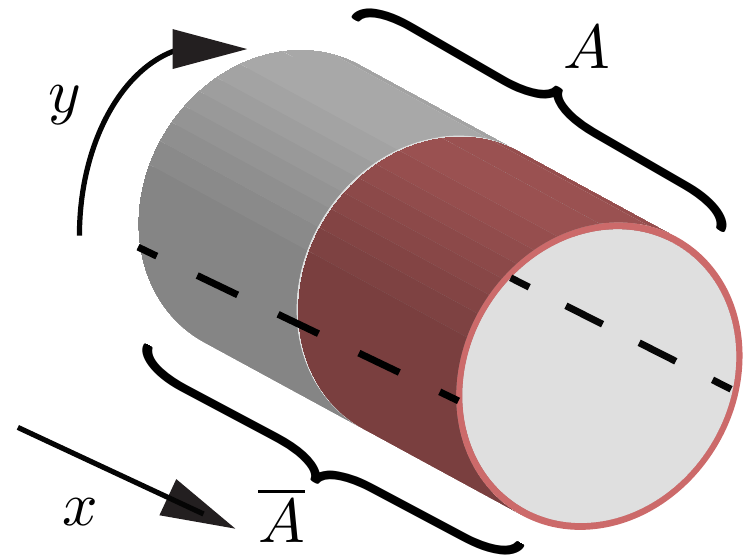}{6}{\small{\textsf{We show the geometry of the system and the entanglement cut. The system is taken on a cylinder with open boundary conditions in the $x$-direction (along the axis of the cylinder) and periodic boundary conditions in the $y$ direction (along the circumference). The entanglement cut is along the $x$ direction and is shown in the dashed lines in the figure. The red region specifies the region on which the velocity parameter $v_y$ is changed with the profile shown in Fig.~\ref{fig:fig3.pdf} in the $x$ direction, but it does not change as a function of $y$.}}}

To observe the change in the entanglement entropy under a boost, we need to change the parameter $v_y$ for the Chern insulator (recall we assume $v_y>0$). When we have open boundary conditions,  and with the cut used in Fig.~\ref{fig:fig2.pdf}, we expect there to be no anomalous change in the parity-odd contribution to the entanglement entropy of the system under a uniform change in  $v_y$ (i.e., a, boost). In this case, since both edges will be boosted identically, then the sum total of the change in the anomalous parity-odd contribution to the entanglement entropy  should vanish. We do note, however, that the total entanglement entropy may still be modified when changing $v_y$ due to, for example, changes of the non-universal parameter $\alpha$ in eq.~\eqref{eq:eeunchanged}. 

Instead, to isolate a single edge as in the domain-wall configuration, we change the velocity parameter $v_y$ near only one of the edges of the Chern insulator lattice system, and have it return to its initial value deep in the bulk, but far from the other edge\footnote{It is sufficient to change it on a region extending into the bulk from the edge that is much larger than the penetration depth of the localized edge states.}. In this case, if there is to be no anomalous entanglement contribution, then the change due to the single edge must be compensated by the boosted bulk. An important difference between the calculations we do in this subsection versus the rest of the paper is that we perform global boosts on the edge (as done in Ref.~\cite{Castro:2014tta}). We focus on such global boosts here to maintain translation invariance in the lattice along the edge direction which simplifies the calculations considerably. The expected answer for the change in the edge-state entanglement entropy under a global boost is $-\frac{c_{L}-c_{R}}{6}\eta$ for a single interval with two endpoints on the edge. This has an extra factor of two compared to  the answer that we found under the local boost prescribed in Fig.~\ref{fig:fig3.pdf} and eq.~\eqref{eq:eedge}.

Before we calculate the bulk-boundary cancellation, let us first show that a lattice chiral fermion itself will have a non-vanishing change in the entanglement entropy when boosted. For free-fermions the correlation matrix method\cite{peschel2004} is a convenient technique that we will use to calculate the change in entanglement entropy of a chiral fermion under such a boost. Now, we could take the full bulk spectrum of our 2+1-d lattice model, project onto the chiral edge modes, and then calculate the entanglement of this subset of states. However, for clarity of presentation, and calculational simplicity, we note that the chiral edge modes are essentially identical to the subset of right-movers (or left-movers) of a 1+1-d single-band tight-binding model of lattice fermions. A hopping model which realizes this analog is
\begin{equation}
H=v_{y}\sum_{j}\left(\frac{1}{2i}c^{\dagger}_{j+1}c_{j}-\frac{1}{2i}c^{\dagger}_{j}c_{j+1}\right)
\end{equation}\noindent which has a dispersion $E(k_y)=v_y\sin k_y.$ This model mimics the lattice chiral fermions except that there are both right and left movers since the dispersion is defined over the entire Brillouin zone. This is a consequence of the Nielsen-Ninomiya theorem. Fortunately, since the model is non-interacting, we can isolate just the left or right movers and calculate their entanglement using this simpler model.

To calculate the entanglement spectrum and entropy we need to construct the correlation matrix. The correlation matrix for this hopping model in 1+1d is given by \cite{peschel2004}
\begin{equation}
\label{eq:cmatrix}
\hat{C}(i,j)=\int_{-\pi}^{\pi} \frac{dq_{y}}{2\pi}e^{iq_{y}(i-j)}n(q_{y})
\end{equation}\noindent where $n(q_{y})$ is the occupation number of the ``edge states" as a function of $q_{y},$ and we have used the matrix element $\langle c^{\dagger}_{i}c_{j}\rangle=e^{iq_y(i-j)}$ . The calculation and end result, for the entanglement entropy is discussed in detail in Appendix~\ref{app:corr_calc}. We find that varying the entropy with respect to the cutoff momentum $k_0$ is $\frac{\delta S}{\delta k_{0}}=-\frac{c_{L}-c_{R}}{6}\times \tan \frac{k_{0}}{2}$. Here $k_0$ is the momentum cut-off for the occupied states in the chiral fermion branch, and it is of interest in this calculation, as shown in Appendix~\ref{app:corr_calc}, since changing the velocity of the chiral fermions acts identically to changing the momentum cut-off.
Interestingly, we find that our result for the boosted entanglement entropy has a lattice factor $\tan \frac{k_{0}}{2}$, which is precisely related to the lattice factor appearing in the Calabrese-Cardy lattice formula for the spatial entanglement entropy of a CFT\cite{Calabrese:2004eu}.  At a cutoff momentum of $k_{0}=\frac{\pi}{2}$, we recover the exact continuum value for a global boost on the edge. When $k_{0}\sim 0,\pi$, we expect the lattice effects to be maximal because we are close to the limit of an empty or fully filled edge band respectively, and the dispersion becomes non-chiral. The fact that we have found lattice modifications to the connection between entanglement and gravity may give hints about how to properly discuss lattice regularizations of gravitational anomalies, although we leave that to future work.

\myfig{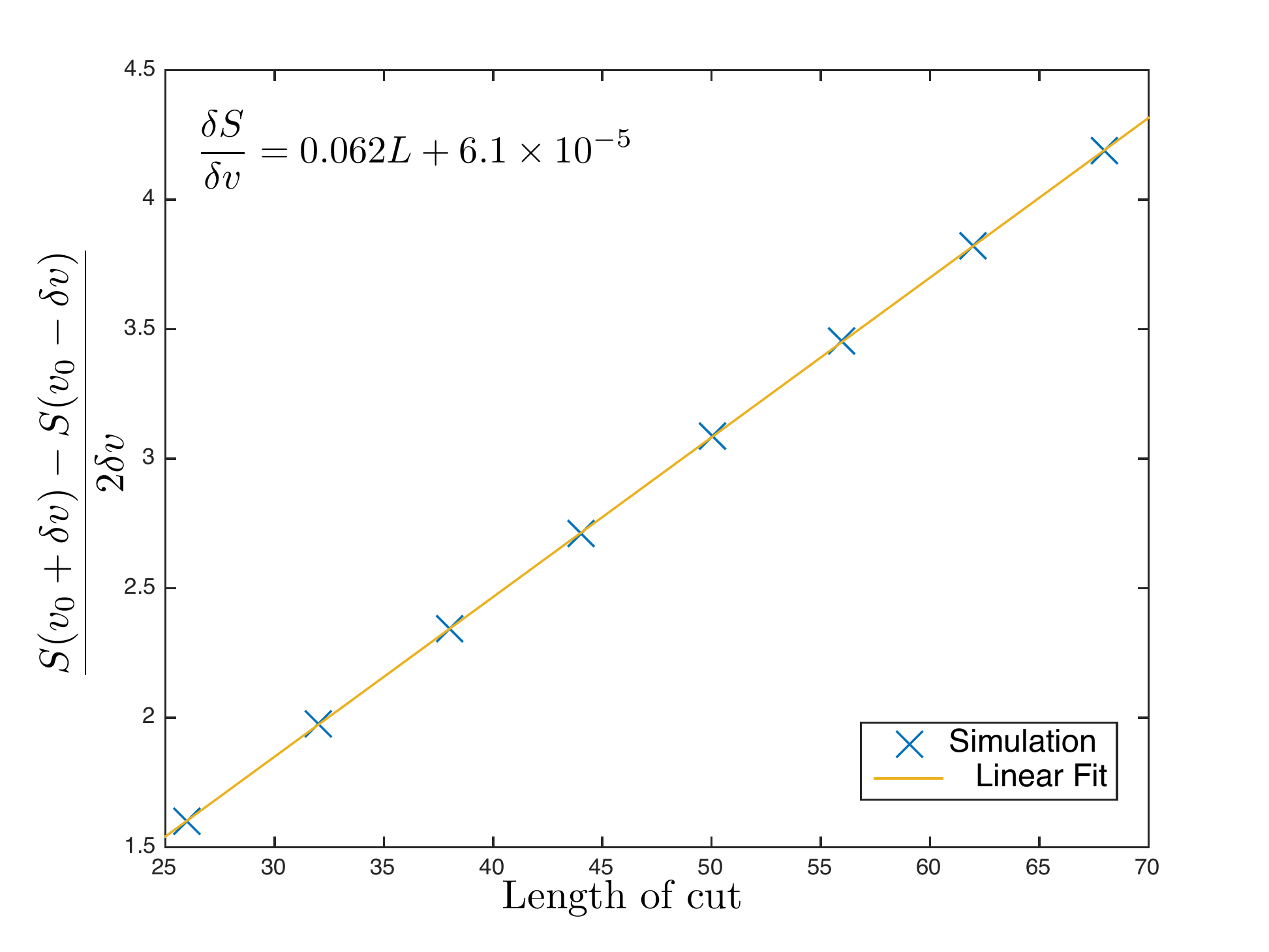}{11}{\small{\textsf{We plot the change in entanglement entropy $\delta S$ vs the length of the entanglement cut (the rectangle including contributions to the length along each edge, and from the sides of the cylinder) of a Chern insulator system with open boundary conditions in the $x$-direction and periodic boundary conditions along the $y$-direction with the entanglement cut from Fig.~\ref{fig:fig11.pdf} and the boost near a left moving edge. The blue cross markers are the simulation results and the yellow line is a linear fit. We notice from the inset fitting equation that the plot is linear in the length of the entanglement cut and the tiny constant offset is likely from finite size errors. The anomalous contribution from a chiral edge state would show up as a finite offset on the linear fit of order ($-\frac{c_{L}-c_{R}}{6}=-\frac{1}{6}$), which we do not see. This indicates cancellation between the bulk and boundary contributions to the entropy. The parameters used in the plot were $v_y=1.0$, $\delta v=0.01$ and $m=1.0$ from the Chern insulator model in eq.~\eqref{eq:CI}.}}}

We have now confirmed that a free chiral fermion gives a modification to the entanglement entropy when boosted. When the entire bulk and boundary system is considered together (with open boundary conditions in $x$ and periodic boundary conditions in $y$ as illustrated in Fig.~\ref{fig:fig11.pdf}), we expect the anomalous change in the parity-odd contribution to the entanglement entropy to cancel out. To extract the parity odd part of the entanglement entropy, our strategy will be to consider the system at positive and negative $v_{y}^{(0)}$ (which correspond to positive and negative Chern number respectively, i.e., opposite parity phases) and change the velocity by a small amount $v_{\pm}=\pm v_y^{(0)}\rightarrow \pm v_y^{(0)}+\delta v$ near one of the edges to isolate the change from the other edge. The parity odd part of the change in entanglement entropy is then calculated as 
\begin{equation}
\label{eq:parityodd}
\frac{\delta S}{\delta v}=\frac{1}{2\delta v}\left\lbrack S(-v_{y}^{(0)}+\delta v)-S(v_{y}^{(0)}+\delta v)\right\rbrack .
\end{equation}\noindent We have seen that a change in entropy due to the boundary chiral fermion should yield a constant contribution to $\frac{\delta S}{\delta v}$ (or equivalently $\frac{\delta S}{\delta k_0}$) which is independent of the size of the entanglement cut. However, what we find in Fig.~\ref{fig:fig12.pdf} is that the total change $\delta S$ is perfectly linear in the length of the entanglement cut, and does not have any signature of a constant offset coming from the chiral edge state. In fact, we can understand the linear-scaling piece as follows. The entanglement entropy from the bulk $S_{bulk}(v_{y}^{(0)})=\alpha(v_{y}^{(0)})L$ is an even function of the velocity $v_{y}^{(0)}$. Thus, in the small $\delta v$ regime, there is an additional non-universal change in eq.~\eqref{eq:parityodd} coming from $\alpha(v_{y}^{(0)})$ that can be approximated as $\frac{\delta S_{bulk}}{\delta v}\approx \alpha'(v_{y}^{(0)})L+\mathcal{O}(\delta v)$. This is where we expect the non-vanishing contribution in Fig. \ref{fig:fig12.pdf} comes from. 

From the linear fitting equation shown in  Fig.~\ref{fig:fig12.pdf} we see that the constant offset is essentially vanishing, and hence  we have demonstrated that the bulk variation does indeed cancel off the universal part of the variation of the boundary entanglement entropy as expected from our earlier calculations. Additional consistency checks were done to check that the $\frac{\delta S}{\delta v}$ plotted in Fig.~\ref{fig:fig12.pdf} is independent of the  $\eta=\delta v$ that was chosen in the small $\eta/v_y^{(0)}$ regime. The result was also found not to depend qualitatively on the the parameter $m$ from the bulk Chern insulator model in eq.~\eqref{eq:CI}, as long as one remains in a gapped topological phase.

\section{Discussion}

In 2+1 dimensional parity violating theories, domain walls between topologically distinct phases support localized 1+1 dimensional gapless excitations. The low energy effective description of these excitations is in terms of (generically) chiral conformal field theories, which possess a variety of anomalous currents. In this paper, we considered the entanglement entropy in such 2+1 dimensional theories, in the situation where the entangling surface intersects the domain wall. We have shown that the anomalous dependence of the entanglement entropy on local boosts of the entangling surface in the 1+1 dimensional theory are accounted for by inflow from the bulk. We have thus termed this phenomenon {\it entanglement inflow}. This is in keeping with the general structure of anomaly inflow, and is a manifestation of the fact that the  1+1 dimensional theory does not really exist in isolation. It is satisfying that entanglement entropy can be consistently defined (i.e., independently of local boosts which preserve the entangling surface) by going to the full bulk description. In a manner of speaking, the bulk theory provides a UV completion of the edge CFT, and it is only natural that one has to go to the bulk to define entanglement entropy in a consistent way.

Interestingly, we were able to give a physical picture for cancellation of the bulk and boundary entanglement responses in terms of (entanglement) energy inflow in the Rindler wedge, mediated by the low-lying chiral states in the entanglement spectrum. We inferred the existence of these states from the requirement of anomaly cancellation, or in other words (background) diffeomorphism invariance, in the Rindler wedge. Turning the logic around, one can deduce the existence of low-lying chiral states in the entanglement spectrum from the physical principle of energy conservation in the Rindler wedge, through thought experiments like the one considered in Fig.~\ref{fig:fig10.pdf}. This gives an interesting new vantage-point to think about the correspondence between the low-lying spectrum of the entanglement Hamiltonian and the spectrum of physical edge states in topological phases. 

We end with some discussion of open problems and future directions. First, it is interesting to consider similar phenomena in higher dimensions \cite{Walltalk}. Gravitational anomalies exist in $4k+2$ dimensions (for $k=1,2\cdots$), and we expect our arguments to carry through in these cases as well. Of course, in order to study these higher dimensional analogues, one presumably has to admit more complicated geometries for the entangling surface. Second, the effect of parity-violation on entanglement entropy  is by itself an interesting arena to explore. Another interesting question is whether the above physics can be described holographically in strongly coupled field theories. Indeed, the anomalous response of the entanglement entropy to boosts was considered in the context of holographic entanglement entropy in \cite{Castro:2014tta}, and more recently in \cite{Guo:2015uqa,Azeyanagi:2015uoa}. In our case, in order to describe a holographic CFT with three spacetime dimensions and a parity domain wall which hosts a lower dimensional chiral CFT, one might consider asymptotically-AdS Janus-like solutions \cite{Bak:2003jk} supported by pseudoscalars \cite{Leigh:2008tt}. It might be an interesting exercise to consider the Ryu-Takayanagi entanglement entropy in this context \cite{Azeyanagi:2007qj}. Since most of the arguments in this paper were based on general principles, there is no reason to suspect any radical modifications at strong coupling. Additionally, it would be interesting to consider the properties of other symmetry-protected topological phases which are not chiral, but still harbor low-energy edge states and entanglement modes.
We leave these questions for future exploration.

\textbf{Note}: While this paper was in its final stages of preparation, references \cite{IqbalWall,Nishioka:2015uka} appeared on the arXiv, which have some overlap with the contents (in particular, section \ref{sec2.2}) of this paper.

\section{Acknowledgements}
We are grateful to Tom Faulkner for enlightening discussions and helpful comments on an earlier draft, and to Aron Wall for correspondence, and for sharing his draft with us a day in advance. TLH and OP would like to thank the Kavli Institute for Theoretical Physics, Santa Barbara (where early work on this project began) for support under the U.S.
National Science Foundation grant number NSF PHY11-25915. TLH and STR were supported by  NSF CAREER DMR-1351895. RGL and OP are supported in part by the U.S. Department of Energy contract DE-FG02-13ER42001.
\section*{Appendix}

\appendix 
\section{Perturbative change in Entanglement Entropy}\label{appA}
In this appendix, we want to prove the formula \eqref{pert1} for the perturbative change in entanglement entropy. Consider a generic quantum field theory that admits a path-integral description in terms of a field $\phi$, which collectively denotes all the fields over which we will integrate. The density matrix corresponding to the ground state wavefunction at $t=0$ is given by
\beq
\rho = \mathrm{lim}_{\beta \to \infty}\;\frac{e^{-\beta \widehat{H}}}{\mathrm{Tr}_{\mathcal{H}}\left(e^{-\beta \widehat{H}}\right)}
\eeq
where $\widehat{H}$ is the Hamiltonian, and $\mathcal{H}$ is the relevant Hilbert space. In the path integral language, we can describe a matrix element of this density matrix as a product of path integrals over the regions $\tau_E>0$ and $\tau_E<0$ of Euclidean space $\re^3$ with the appropriate boundary conditions
\beq
\langle \phi_-|\rho|\phi_+\rangle = \frac{1}{Z}\int_{\phi(0^-,\bdx)= \phi_-(\bdx)}[D\phi]_{\tau_E<0}\;e^{-S[\phi]}\int_{\phi(0^+,\bdx)= \phi_+(\bdx)}[D\phi]_{\tau_E>0}\;e^{-S[\phi]}
\eeq
where we have denoted the spatial coordinates collectively as $\bdx=(y,x)$, and $Z$ is the partition function of the theory on $\re^3$. 

Let us now denote the reduced density matrix for the half space $A=\{\bdx^A=(y,x)|y>0\}$ by $\rho_A$. The matrix element $\langle \phi^A_-|\rho_A|\phi^A_+\rangle$ is then given by gluing the above path integrals along the complementary space $\bar{A}$ at $\tau_E=0$ 
\beq
\langle \phi^A_-|\rho_A|\phi^A_+\rangle=\frac{1}{Z}\int^{\phi(0^-,\bdx^A)=\phi^A_-}_{\phi(0^+,\bdx^A)=\phi^A_+}[D\phi]\;e^{-S[\phi]}
\eeq
where $\bdx^A$ are spatial coordinates on $A$, and $\phi^A_{\pm}$ denotes a field configuration on $A$. By slicing this path integral along the angular direction $\varphi$ in the $(\tau_E,y)$ plane, it becomes immediately clear that the reduced density matrix can be written in operator form as \cite{Bisognano:1976za}
\beq\label{EH}
\rho_A = \frac{e^{-2\pi\widehat{K}}}{\mathrm{Tr}_{\mathcal{H}_A}\left(e^{-2\pi\widehat{K}}\right)}
\eeq
where $\widehat{K}$ is the generator of $\varphi$-rotations 
\beq
\hK = \int_{-\infty}^{\infty} dx\int_0^{\infty}dy\;y\;\widehat{T}^{00}(0,y,x).
\eeq
Equation \eqref{EH} shows that, up to an overall shift coming from the normalization, the entanglement Hamiltonian in this case is given by 
\beq
\widehat{H}_E = -\mathrm{ln}\;\rho_A=2\pi\widehat{K}+\mathrm{ln}\;\mathrm{Tr}_{\mathcal{H}_A}\left(e^{-2\pi\widehat{K}}\right).
\eeq

Now, if we turn on a small (background) metric deformation $\delta g_{\mu\nu}$ we find
\beq
\langle \phi^A_-|\left(\rho_A+\delta \rho_A\right)|\phi^A_+\rangle=\frac{1}{(Z+\delta Z)}\int^{\phi(0^-,\bdx^A)=\phi^A_-}_{\phi(0^+,\bdx^A)=\phi^A_+}[D\phi]\;e^{-S[\phi]+\frac{1}{2}\int_{\re^3} d\tau_E d^{d-1}\bdx\;\delta g^{\mu\nu}(\tau_E,\bdx)\hT_{\mu\nu}(\tau_E,\bdx)+\cdots}\label{dm1}
\eeq
and therefore to linear order in $\delta g$, we have
\beq
\delta \rho_A = \frac{1}{2Z}\int_{\phi^R_+}^{\phi_-^R}[D\phi]\;e^{-S[\phi]}\int_{\re^3} d\tau_E d^{d-1}\bdx\; \delta g^{\mu\nu}(\tau_E,\bdx)\left(\hT_{\mu\nu}(\tau_E,\bdx)-\langle \hT_{\mu\nu}(\tau_E,\bdx)\rangle\right).
\eeq
From equation \eqref{EH}, we can then infer the following operator expression for the change in the reduced density matrix
\beq
\delta\rho_A = \frac{1}{2}\int_{\re^3} d\tau_E d^{d-1}\bdx\;\delta g^{\mu\nu}(\tau_E,\bdx)\Big(\rho_A\hT_{\mu\nu}(\tau,\bdx)-\rho_A\langle\hT_{\mu\nu}(\tau,\bdx)\rangle\Big)
\eeq
The operator $\hT(\tau_E,\bdx)$ above is to be interpreted (from the point of view of the reduced density matrix) as a Heisenberg operator
\beq
\hT^{\mu\nu}(\varphi,r,x) = {\left(R^{-1}(\varphi)\right)^{\mu}}_{\lambda}{\left(R^{-1}(\varphi)\right)^{\nu}}_{\sigma}e^{\varphi \hK}\hT^{\lambda\sigma}(0, r, x)e^{-\varphi \hK}
\eeq
where $(\varphi,r)$ are polar coordinates in the $(\tau_E,y)$ plane, and $R(\varphi)$ is the appropriate rotation matrix. 

Finally, from the definition of entanglement entropy, we have
\beq
\delta S_{EE}=\mathrm{Tr}_{\mathcal{H}_A}\left(\delta\rho_A\;\widehat{H}_E\right)
\eeq
Putting everything together, we find
\beqn
\delta S_{EE}&=&\frac{1}{2}\int_{\re^3} d\tau_E d^{d-1}\bdx\;\delta g^{\mu\nu}(\tau_E,\bdx)\mathrm{Tr}_{\mathcal{H}_A}\Big(\rho_A\hT_{\mu\nu}(\tau,\bdx)\widehat{H}_E-\rho_A\langle\hT_{\mu\nu}(\tau,\bdx)\rangle\widehat{H}_E\Big)\nonumber\\
&=&\frac{1}{2}\int_{\re^3} d\tau_E d^{d-1}\bdx\;\delta g^{\mu\nu}(\tau_E,\bdx)\Big(\left\langle \hT_{\mu\nu}(\tau,\bdx)\widehat{H}_E\right\rangle-\left\langle\hT_{\mu\nu}(\tau,\bdx)\right\rangle\left\langle \widehat{H}_E\right\rangle\Big)\nonumber\\
&=&\frac{1}{2}\int_{\re^3} d\tau_E d^{d-1}\bdx\;\delta g^{\mu\nu}(\tau_E,\bdx)\left\langle \hT_{\mu\nu}(\tau,\bdx)\widehat{H}_E\right\rangle_{\mathrm{conn.}}
\eeqn
where the correlation functions appearing above are Euclidean correlation functions on $\re^3$, and in the last line we have used the definition of a connected correlation function. The above integral is divergent in general, and one must be careful to regulate it by removing a tubular neighbourhood around the point $\tau_E=y=0$
\beq
\delta S_{EE} = \frac{1}{2}\int_{\tau_E^2+y^2\geq a^2} d\tau_E d^{d-1}\bdx\;\delta g^{\mu\nu}(\tau_E,\bdx)\left\langle \hT_{\mu\nu}(\tau,\bdx)\widehat{H}_E\right\rangle_{\mathrm{conn.}}
\eeq
Thus, we arrive at the expression used in the main text.

\section{Entanglement spectrum for free Dirac fermions}\label{appES}
In this section, we demonstrate explicitly that in the case of a free $d=2+1$ massive Dirac fermion with a mass domain wall, there are chiral modes along the domain wall, and more importantly, along the entangling surface. We will work in Lorentzian signature in this section, because the spectrum of the entanglement Hamiltonian will be much simpler to interpret in this case. In order to get the spectrum, we simply write down the Dirac equation on the Rindler wedge in $\re^{1,2}$ with a domain wall profile for the mass $m(x)$. We then conformally map to $\re\times \mathbb{H}^2$ (this is not necessary, but we do so to make contact with the discussion in the main text), and solve the resulting Dirac equation on $\re\times \mathbb{H}^2$.  

The Dirac equation on a general background is given by
\beq
\left(\gamma^a\ue_a^{\mu}D_{\mu}+m\right)\Psi = \Big(\gamma^{a}\ue^{\mu}_a\pa_{\mu}+\frac{1}{4}\ue^{\mu}_a\omega_{\mu;bc}\gamma^{a}\gamma^{bc}+m(x)\Big)\Psi = 0
\eeq
where $\ue_a$ is an orthonormal frame on spacetime, $\omega^a{}_b$ is the spin-connection, and $\gamma^a$ are matrices satisfying the Clifford algebra $\{\gamma^a,\gamma^b\} =2\eta^{ab}\boldsymbol{1}$. A simple property of the Dirac equation is that it is covariant under the transformation \cite{Nieh:1981xk}
\beq
\tilde{e}^a = \Omega e^a,\;\;\tilde{\Psi} = \Omega^{-\frac{d-1}{2}}\Psi,\;\;\tilde{m}=\Omega^{-1}m
\eeq
where $\Omega$ is an arbitrary function on spacetime and $d$ is the spacetime dimension. We can use this fact together with the conformal equivalence of  the Rindler wedge with $\re^1\times \mathbb{H}^2$ to solve the Dirac equation on the latter geometry with the mass $\tilde{m} = \Omega^{-1}m$. Let us parametrize $\re\times \mathbb{H}^2$ with the coordinates $(T,z,x)$ with $T$ being the time. It is convenient to use the following coframe 
\beq
\tilde{e}^0 = dT,\;\; \tilde{e}^1=\frac{dz}{z},\;\;\tilde{e}^2 = \frac{dx}{z}.
\eeq
The (Levi-Civita) spin connection in this case is given by
\beq
\tilde{\omega}_{12} = \tilde{e}^2=\frac{dx}{z},\;\; \tilde{\omega}_{01} = \tilde{\omega}_{02} = 0
\eeq
and the mass is given by $\tilde{m} = z m(x)$. The Dirac equation on $\re^1\times \mathbb{H}^2$ is (dropping the tildes for convenience)
\beq
\Big(\gamma^0\pa_{T}+\gamma^1(z\pa_z-\tfrac{1}{2})+\gamma^2z\pa_x+zm(x)\Big)\Psi = 0.
\eeq 
We're interested in the situation where $m(x)$ has a domain wall profile, which means that $m(x)$ interpolates between $-m_0$ and $+m_0$ as $x$ runs from $-\infty$ to $+\infty$, $m_0$ being a positive constant. For simplicity, we will take $m(x)=m_0\;\mathrm{sign}(x)$.

In the presence of the above mass domain wall profile, one finds a special normalizable solution to the Dirac equation of the form
\beq
\Psi^{(k)}_{DW}(T,z,x)= e^{-mx} \sqrt{|m|z}e^{-i\omega(k)T-ik\ln z}\psi_{0}(k)
\eeq
\beq
\omega(k)=k,\;\;\gamma^2\psi_0(k)= + \psi_0(k)
\eeq
This is a zero-mode of the Dirac operator which is localized on the domain wall between the two-phases $m<0$ and $m>0$ (hence the subscript $DW$), and is chiral with respect to the domain wall chirality matrix $\gamma^2$. In the limit $m\to\infty$ (i.e., at energy scales small compared to the mass), we can model the dynamics of this mode as a $1+1$-dimensional chiral fermion.

For $|x|>> 0$ (i.e., far away from the domain wall), it is a simpler matter to construct generic solutions. Going to Fourier space in $T$ and $x$
\beq
\Psi(T,z,x) = \int_{-\infty}^{\infty}\frac{dk}{2\pi}\int_{-\infty}^{\infty}\frac{d\omega}{2\pi}e^{ikx+i\omega T}\Psi(\omega,k;z)
\eeq
and writing the Dirac equation in terms of $\psi_{(\pm)}=\frac{1}{2}(1\pm \gamma^2)\Psi$, we obtain
\beq
iz\pa_z\psi_{(+)}(\omega,k;z)-\left(\omega+\tfrac{i}{2}\right) \psi_{(+)}(\omega,k;z)=z(ik-m)\psi_{(-)}(\omega,k;z)
\eeq
\beq
iz\pa_z\psi_{(-)}(\omega,k;z)+\left(\omega-\tfrac{i}{2}\right) \psi_{(-)}(\omega,k;z)=z(ik+m)\psi_{(+)}(\omega,k;z)
\eeq
Eliminating $\psi_{(-)}$, and defining $\psi_{(+)}=z \hat{\psi}_{(+)}$ we obtain
\beq
z^2\pa_z^2\hat{\psi}_{(+)}(\omega,k;z)+z\pa_z\hat{\psi}_{(+)}(\omega,k;z)-\left[\left(\tfrac{1}{2}+i\omega\right)^2+z^2(m^2+k^2)\right]\hat{\psi}_{(+)}(\omega,k;z)=0.
\eeq
Therefore, generic bulk solutions can be written in terms of modified Bessel functions\footnote{We have dropped solutions involving $I_{\pm \frac{1}{2}+i\omega}\left(\sqrt{k^2+m^2}z\right)$ because they diverge for large $z$.}
\beq
\psi^{(\omega,k)}_{(+)}(T,z,x)=z\;K_{\frac{1}{2}+i\omega}\left(\sqrt{k^2+m^2}z\right)e^{i\omega T+ikx}\;\chi_0(\omega,k)
\eeq
\beq
\psi^{(\omega,k)}_{(-)}(T,z,x)=\frac{-k+im}{\sqrt{k^2+m^2}}z\;K_{-\frac{1}{2}+i\omega}\left(\sqrt{k^2+m^2}z\right)e^{i\omega T+ikx}\;\chi_0(\omega,k)
\eeq
which are oscillatory for small $z$ and exponentially decay for large $z$. To identify modes localized on the entangling surface, care must be taken to regulate the theory. Here, we choose to cut off the Rindler wedge at $z=a$, with ``brick-wall'' boundary conditions \cite{'tHooft1985727} at $z=a$
\beq
\gamma^1\Psi(T,a,x) = \sigma \Psi(T,a,x)
\eeq
where $\sigma=\pm 1$. We will only construct solutions for the region $z \geq a$, and the above boundary condition ensures that there is no flow of charge or energy across the cutoff surface. The boundary condition implies
\beq
K_{\frac{1}{2}+i\omega}\left(a\sqrt{k^2+m^2}\right)=\sigma\frac{m+ik}{\sqrt{k^2+m^2}}\;K_{-\frac{1}{2}+i\omega}\left(a\sqrt{k^2+m^2}\right).
\eeq
\myfig{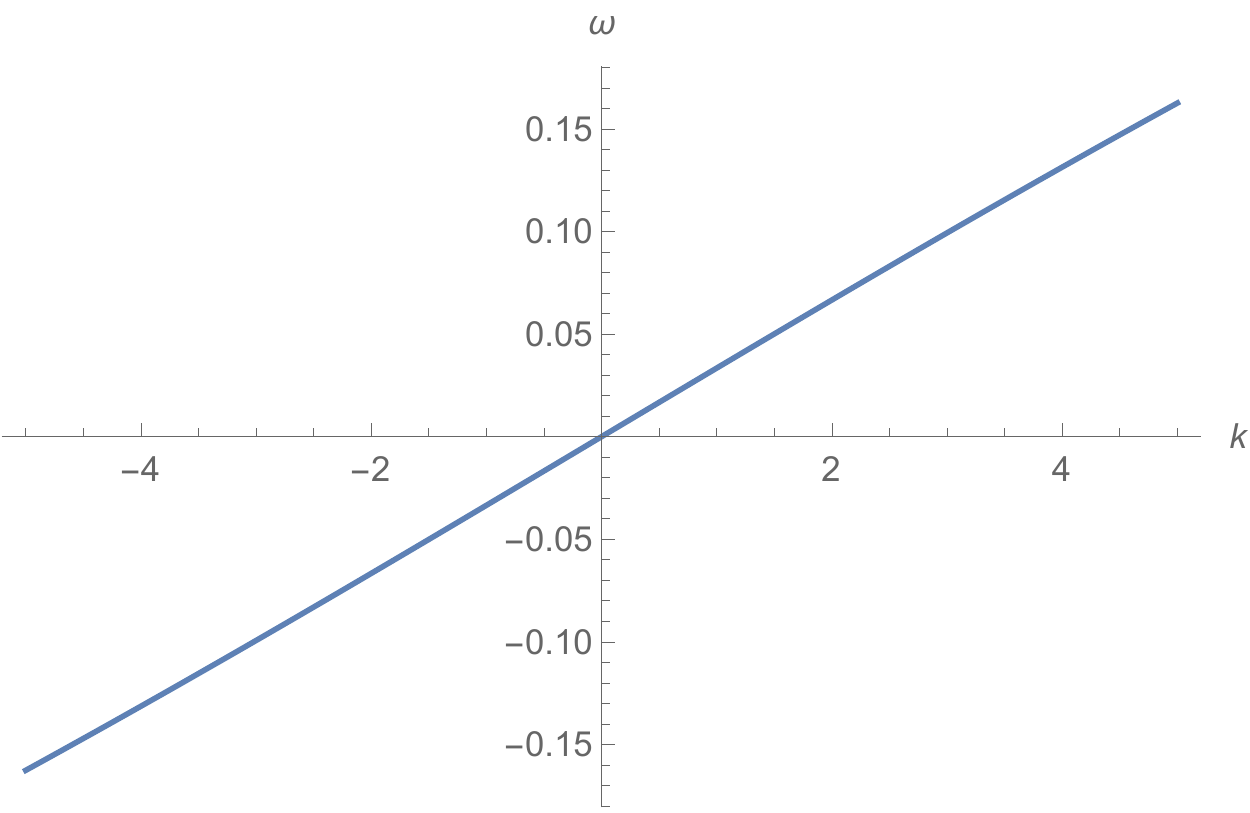}{8}{\small{\textsf{The entanglement spectrum corresponding to the boundary condition with $\sigma=\mathrm{sign}(m)$ for $m=10$, and $a=0.01$.}}}
This equation can be solved numerically to obtain $\omega$ as a function of $k$, namely the entanglement spectrum. For $\sigma=\mathrm{sign}(m)$, we find spectrum (see Fig.~\ref{fig:fig13.pdf})
\beq
\omega(k) = v k+\cdots
\eeq
where the ellipses denote terms of order $\left(\frac{k}{m}\right)^3$. The velocity $v$ in this case is non-universal, and is given by 
\beq
v=\frac{e^{-2ma}}{2m\Gamma(0,2ma)}.
\eeq
For $|k|$ small compared to $|m|$, the wavefunctions are localized near the regularized entangling surface $z=a$, and are chiral with respect to the entangling surface's chirality matrix $\gamma^1$, up to $O(k/m)$ corrections. Thus, we see that the cutoff entangling surface hosts a chiral mode. Quantizing this mode gives us Casimir momentum density on the entangling surface, which is what was used in the main text.

The other branch $\sigma=-\mathrm{sign}(m)$ gives rise to gapped states in the entanglement spectrum. These states are oscillatory for $z$ large compared to $a$, and should be interpreted as bulk states. Since the entanglement entropy is dominated by the low-lying spectrum of the entanglement Hamiltonian, it suffices to focus on the low (entanglement) energy states, which in the present case consist of the chiral modes localized on the domain wall and the regulated entangling surface. 

\section{Correlation matrix and entanglement entropy for an edge chiral fermion}\label{app:corr_calc}
The correlation matrix as given in eq.~\eqref{eq:cmatrix} is 
\begin{equation}
\hat{C}(i,j)=\int_{-\pi}^{\pi} \frac{dq_{y}}{2\pi}e^{i(q_{y})(i-j)}n(q_{y})
\end{equation}\noindent where $n(q)$ is the occupation number as a function of momentum. For the case of a right moving edge state with the dispersion $E=v_{y}\sin q_{y}$ which, for a single chiral fermion does not traverse the entirety of the edge BZ, $n(q_{y})=1$ for $-k_{0}\leq q_{y}\leq 0$ (assuming $k_{0}>0$) where $k_{0}$ is a cutoff determined from where the edge band $v_{y}\sin q_{y}$ meets the bulk band $E_{-}$. We can thus simplify the correlation matrix to 
\begin{equation}
\label{eq:corr_CIapp}
\hat{C}_{R}(i,j)=\frac{1-e^{-ik_{0}(i-j)}}{2\pi i(i-j)}=e^{-ik_{0}(m-n)/2}\frac{\sin k_{0}(m-n)/2}{\pi(m-n)}.
\end{equation} The eigenvalues of this matrix when $i,j$ are restricted to a subsystem can be used to calculate the entanglement entropy. To move from a right moving edge to a left moving edge, we simply need to flip $k_{0}\rightarrow -k_{0}$ in eq.~\eqref{eq:corr_CIapp}. This corresponds to choosing $n(q_{y})=1$ for $0\leq q_{y}\leq k_{0}$. 

The phase factor from \eqref{eq:corr_CIapp} does not affect the eigenvalues of the correlation matrix, which we can see as follows. The eigenvalue equation for the correlation matrix is given by $\sum_{j}C_{ij}\phi_{k}(j)=\zeta_{k}\phi_{k}(i)$. Hence, the phase factor in $C_{ij}$ can be absorbed into the $\phi_{k}(i),\phi_{k}(j)$'s so that $\zeta_{k}$ remain the same. Modulo the phase factor, the correlation matrix from this problem is identical to what is found for a 1D tight-binding electron, but with a Fermi wavevector of  $k_{0}/2$~\cite{EislerPeschel}. The entanglement entropy is given by 
\beq
\label{eq:ee_sum}
S=\sum_{k}\left[\log(1+e^{-\epsilon_{k}})+\frac{\epsilon_{k}}{1+e^{\epsilon_{k}}}\right]
\eeq\noindent where $\epsilon_{k}$ are the eigenvalues of the entanglement Hamiltonian and are related to $\zeta_{k}$ as $\zeta_{k}=\frac{1}{e^{\epsilon_{k}}+1}$. Quoting from Refs. \cite{EislerPeschel,peschel2004}, we have that $\frac{d\vert\epsilon\vert}{dk}=\frac{\pi^{2}}{\log 2L\sin k_{0}/2}$ where $L$ is the length of the subsystem in question. This can be used to perform the integration in eq.~\eqref{eq:ee_sum}, $S=\frac{2\log 2L\sin k_{0}/2}{\pi^{2}}\int_{0}^{\infty} d\epsilon\left[\log(1+e^{-\epsilon_{k}})+\frac{\epsilon_{k}}{1+e^{\epsilon_{k}}}\right]$ where each term in the integration gives us a factor of $\frac{\pi^{2}}{12}$. There is an extra factor of two overall which we have inserted when converting from the sum to the integral to account for both positive and negative $\epsilon_{k}$. The end result is that $S=\frac{\log (2L\sin k_{0}/2)}{3}$. Under a variation of the cutoff $k_{0}$, we see that 
\beq 
\frac{\delta S}{\delta k_{0}}=\tan \frac{k_{0}}{2}\times \frac{1}{6}
\eeq\noindent for a right moving edge. So, in general we have $\frac{\delta S}{\delta k_{0}}=-\tan \frac{k_{0}}{2}\times \frac{c_{L}-c_{R}}{6}$.


\begin{thebibliography}{10}

\bibitem{Calabrese:2004eu}
P.~Calabrese and J.~L. Cardy, ``{Entanglement entropy and quantum field
  theory},'' \href{http://dx.doi.org/10.1088/1742-5468/2004/06/P06002}{{\em J.
  Stat. Mech.} {\bf 0406} (2004)  P06002},
\href{http://arxiv.org/abs/hep-th/0405152}{{\tt arXiv:hep-th/0405152
  [hep-th]}}.

\bibitem{1751-8121-42-50-504005}
P.~Calabrese and J.~Cardy, ``Entanglement entropy and conformal field theory,''
  {\em Journal of Physics A: Mathematical and Theoretical} {\bf 42} (2009)
  no.~50, 504005. \url{http://stacks.iop.org/1751-8121/42/i=50/a=504005}.

\bibitem{Casini:2009sr}
H.~Casini and M.~Huerta, ``{Entanglement entropy in free quantum field
  theory},'' \href{http://dx.doi.org/10.1088/1751-8113/42/50/504007}{{\em J.
  Phys.} {\bf A42} (2009)  504007},
\href{http://arxiv.org/abs/0905.2562}{{\tt arXiv:0905.2562 [hep-th]}}.

\bibitem{hammakitaev}
A.~Hamma, R.~Ionicioiu, and P.~Zanardi, ``Ground state entanglement and
  geometric entropy in the kitaev model,'' {\em Physics Letters A} {\bf 337}
  (2005) no.~1, 22--28.

\bibitem{Kitaev:2005dm}
A.~Kitaev and J.~Preskill, ``{Topological entanglement entropy},''
  \href{http://dx.doi.org/10.1103/PhysRevLett.96.110404}{{\em Phys. Rev. Lett.}
  {\bf 96} (2006)  110404},
\href{http://arxiv.org/abs/hep-th/0510092}{{\tt arXiv:hep-th/0510092
  [hep-th]}}.

\bibitem{PhysRevLett.96.110405}
M.~Levin and X.-G. Wen,
  \href{http://dx.doi.org/10.1103/PhysRevLett.96.110405}{``Detecting
  topological order in a ground state wave function,''{\em Phys. Rev. Lett.}
  {\bf 96} (Mar, 2006)  110405}.
  \url{http://link.aps.org/doi/10.1103/PhysRevLett.96.110405}.

\bibitem{Dong:2008ft}
S.~Dong, E.~Fradkin, R.~G. Leigh, and S.~Nowling, ``{Topological Entanglement
  Entropy in Chern-Simons Theories and Quantum Hall Fluids},''
  \href{http://dx.doi.org/10.1088/1126-6708/2008/05/016}{{\em JHEP} {\bf 05}
  (2008)  016},
\href{http://arxiv.org/abs/0802.3231}{{\tt arXiv:0802.3231 [hep-th]}}.

\bibitem{lihaldane}
H.~Li and F.~D.~M. Haldane, ``Entanglement spectrum as a generalization of
  entanglement entropy: Identification of topological order in non-abelian
  fractional quantum hall effect states,'' {\em Phys. Rev. Lett.} {\bf 101}
  (2008) no.~1, 010504.

\bibitem{Flammia2009}
S.~T. Flammia, A.~Hamma, T.~L. Hughes, and X.-G. Wen,
  \href{http://dx.doi.org/10.1103/PhysRevLett.103.261601}{``Topological
  entanglement r\'enyi entropy and reduced density matrix structure,''{\em
  Phys. Rev. Lett.} {\bf 103} (Dec, 2009)  261601}.

\bibitem{Regnault2009}
N.~Regnault, B.~A. Bernevig, and F.~D.~M. Haldane,
  \href{http://dx.doi.org/10.1103/PhysRevLett.103.016801}{``Topological
  entanglement and clustering of jain hierarchy states,''{\em Phys. Rev. Lett.}
  {\bf 103} (Jun, 2009)  016801}.

\bibitem{Yao2010}
H.~Yao and X.-L. Qi,
  \href{http://dx.doi.org/10.1103/PhysRevLett.105.080501}{``Entanglement
  entropy and entanglement spectrum of the kitaev model,''{\em Phys. Rev.
  Lett.} {\bf 105} (Aug, 2010)  080501}.

\bibitem{sterdyniak2011a}
A.~Sterdyniak, B.~Bernevig, N.~Regnault, and F.~Haldane, ``The hierarchical
  structure in the orbital entanglement spectrum of fractional quantum hall
  systems,'' {\em New Journal of Physics} {\bf 13} (2011) no.~10, 105001.

\bibitem{hermanns2011}
M.~Hermanns, A.~Chandran, N.~Regnault, and B.~A. Bernevig, ``Haldane statistics
  in the finite-size entanglement spectra of 1/m fractional quantum hall
  states,'' {\em Physical Review B} {\bf 84} (2011) no.~12, 121309.

\bibitem{Ryu:2006bv}
S.~Ryu and T.~Takayanagi, ``{Holographic derivation of entanglement entropy
  from AdS/CFT},'' \href{http://dx.doi.org/10.1103/PhysRevLett.96.181602}{{\em
  Phys. Rev. Lett.} {\bf 96} (2006)  181602},
\href{http://arxiv.org/abs/hep-th/0603001}{{\tt arXiv:hep-th/0603001
  [hep-th]}}.

\bibitem{Ryu:2006ef}
S.~Ryu and T.~Takayanagi, ``{Aspects of Holographic Entanglement Entropy},''
  \href{http://dx.doi.org/10.1088/1126-6708/2006/08/045}{{\em JHEP} {\bf 08}
  (2006)  045},
\href{http://arxiv.org/abs/hep-th/0605073}{{\tt arXiv:hep-th/0605073
  [hep-th]}}.

\bibitem{VanRaamsdonk:2010pw}
M.~Van~Raamsdonk, ``{Building up spacetime with quantum entanglement},''
  \href{http://dx.doi.org/10.1007/s10714-010-1034-0,
  10.1142/S0218271810018529}{{\em Gen. Rel. Grav.} {\bf 42} (2010)
  2323--2329}, \href{http://arxiv.org/abs/1005.3035}{{\tt arXiv:1005.3035
  [hep-th]}}.
[Int. J. Mod. Phys.D19,2429(2010)].

\bibitem{Faulkner:2013ica}
T.~Faulkner, M.~Guica, T.~Hartman, R.~C. Myers, and M.~Van~Raamsdonk,
  ``{Gravitation from Entanglement in Holographic CFTs},''
  \href{http://dx.doi.org/10.1007/JHEP03(2014)051}{{\em JHEP} {\bf 03} (2014)
  051},
\href{http://arxiv.org/abs/1312.7856}{{\tt arXiv:1312.7856 [hep-th]}}.

\bibitem{Headrick:2014cta}
M.~Headrick, V.~E. Hubeny, A.~Lawrence, and M.~Rangamani, ``{Causality \&
  holographic entanglement entropy},''
  \href{http://dx.doi.org/10.1007/JHEP12(2014)162}{{\em JHEP} {\bf 12} (2014)
  162},
\href{http://arxiv.org/abs/1408.6300}{{\tt arXiv:1408.6300 [hep-th]}}.

\bibitem{AlvarezGaume:1983ig}
L.~Alvarez-Gaume and E.~Witten, ``{Gravitational Anomalies},''
\href{http://dx.doi.org/10.1016/0550-3213(84)90066-X}{{\em Nucl. Phys.} {\bf
  B234} (1984)  269}.

\bibitem{AlvarezGaume:1984dr}
L.~Alvarez-Gaume and P.~H. Ginsparg, ``{The Structure of Gauge and
  Gravitational Anomalies},''
  \href{http://dx.doi.org/10.1016/0003-4916(85)90087-9}{{\em Annals Phys.} {\bf
  161} (1985)  423}.
[Erratum: Annals Phys.171,233(1986)].

\bibitem{Callan:1984sa}
C.~G. Callan, Jr. and J.~A. Harvey, ``{Anomalies and Fermion Zero Modes on
  Strings and Domain Walls},''
\href{http://dx.doi.org/10.1016/0550-3213(85)90489-4}{{\em Nucl. Phys.} {\bf
  B250} (1985)  427}.

\bibitem{AlvarezGaume:1984nf}
L.~Alvarez-Gaume, S.~Della~Pietra, and G.~W. Moore, ``{Anomalies and Odd
  Dimensions},''
\href{http://dx.doi.org/10.1016/0003-4916(85)90383-5}{{\em Annals Phys.} {\bf
  163} (1985)  288}.

\bibitem{Parrikar:2014usa}
O.~Parrikar, T.~L. Hughes, and R.~G. Leigh, ``{Torsion, Parity-odd Response and
  Anomalies in Topological States},''
  \href{http://dx.doi.org/10.1103/PhysRevD.90.105004}{{\em Phys. Rev.} {\bf
  D90} (2014) no.~10, 105004},
\href{http://arxiv.org/abs/1407.7043}{{\tt arXiv:1407.7043
  [cond-mat.mes-hall]}}.

\bibitem{haldane1988}
F.~D.~M. Haldane, ``Model for a quantum hall effect without landau levels:
  Condensed-matter realization of the" parity anomaly",'' {\em Phys. Rev.
  Lett.} {\bf 61} (1988) no.~18, 2015.

\bibitem{wenreview}
X.~G. Wen {\em Adv. Phys.} {\bf 44} (1995)  405.

\bibitem{Witten:1988hf}
E.~Witten, ``{Quantum Field Theory and the Jones Polynomial},''
\href{http://dx.doi.org/10.1007/BF01217730}{{\em Commun. Math. Phys.} {\bf 121}
  (1989)  351--399}.

\bibitem{Perlmutter:2013paa}
E.~Perlmutter, ``{Comments on Renyi entropy in AdS$_3$/CFT$_2$},''
  \href{http://dx.doi.org/10.1007/JHEP05(2014)052}{{\em JHEP} {\bf 05} (2014)
  052},
\href{http://arxiv.org/abs/1312.5740}{{\tt arXiv:1312.5740 [hep-th]}}.

\bibitem{Castro:2014tta}
A.~Castro, S.~Detournay, N.~Iqbal, and E.~Perlmutter, ``{Holographic
  entanglement entropy and gravitational anomalies},''
  \href{http://dx.doi.org/10.1007/JHEP07(2014)114}{{\em JHEP} {\bf 07} (2014)
  114},
\href{http://arxiv.org/abs/1405.2792}{{\tt arXiv:1405.2792 [hep-th]}}.

\bibitem{Walltalk}
A.~C. Wall, ``Frame anomalies for entanglement in chiral theories.'' Talk given
  at kitp -- \url{http://online.kitp.ucsb.edu/online/entangled15/wall/}.

\bibitem{IqbalWall}
N.~Iqbal and A.~C. Wall, ``{Anomalies of the Entanglement Entropy in Chiral
  Theories},''
\href{http://arxiv.org/abs/1509.04325}{{\tt arXiv:1509.04325 [hep-th]}}.

\bibitem{Nishiokatalk}
T.~Nishioka, ``Anomalies and entanglement entropy.'' Talk given at strings 2015
  -- \url{https://strings2015.icts.res.in/speakerProfile.php?sId=42}.

\bibitem{Nishioka:2015uka}
T.~Nishioka and A.~Yarom, ``{Anomalies and Entanglement Entropy},''
\href{http://arxiv.org/abs/1509.04288}{{\tt arXiv:1509.04288 [hep-th]}}.

\bibitem{Wall:2011kb}
A.~C. Wall, ``{Testing the Generalized Second Law in 1+1 dimensional Conformal
  Vacua: An Argument for the Causal Horizon},''
  \href{http://dx.doi.org/10.1103/PhysRevD.85.024015}{{\em Phys. Rev.} {\bf
  D85} (2012)  024015},
\href{http://arxiv.org/abs/1105.3520}{{\tt arXiv:1105.3520 [gr-qc]}}.

\bibitem{Rosenhaus:2014woa}
V.~Rosenhaus and M.~Smolkin, ``{Entanglement Entropy: A Perturbative
  Calculation},'' \href{http://dx.doi.org/10.1007/JHEP12(2014)179}{{\em JHEP}
  {\bf 12} (2014)  179},
\href{http://arxiv.org/abs/1403.3733}{{\tt arXiv:1403.3733 [hep-th]}}.

\bibitem{Rosenhaus:2014zza}
V.~Rosenhaus and M.~Smolkin, ``{Entanglement Entropy for Relevant and Geometric
  Perturbations},'' \href{http://dx.doi.org/10.1007/JHEP02(2015)015}{{\em JHEP}
  {\bf 02} (2015)  015},
\href{http://arxiv.org/abs/1410.6530}{{\tt arXiv:1410.6530 [hep-th]}}.

\bibitem{Allais:2014ata}
A.~Allais and M.~Mezei, ``{Some results on the shape dependence of entanglement
  and Rényi entropies},''
  \href{http://dx.doi.org/10.1103/PhysRevD.91.046002}{{\em Phys. Rev.} {\bf
  D91} (2015) no.~4, 046002},
\href{http://arxiv.org/abs/1407.7249}{{\tt arXiv:1407.7249 [hep-th]}}.

\bibitem{Faulkner:2014jva}
T.~Faulkner, ``{Bulk Emergence and the RG Flow of Entanglement Entropy},''
  \href{http://dx.doi.org/10.1007/JHEP05(2015)033}{{\em JHEP} {\bf 05} (2015)
  033},
\href{http://arxiv.org/abs/1412.5648}{{\tt arXiv:1412.5648 [hep-th]}}.

\bibitem{Leigh:2003ez}
R.~G. Leigh and A.~C. Petkou, ``{SL(2,Z) action on three-dimensional CFTs and
  holography},'' \href{http://dx.doi.org/10.1088/1126-6708/2003/12/020}{{\em
  JHEP} {\bf 12} (2003)  020},
\href{http://arxiv.org/abs/hep-th/0309177}{{\tt arXiv:hep-th/0309177
  [hep-th]}}.

\bibitem{Closset:2012vp}
C.~Closset, T.~T. Dumitrescu, G.~Festuccia, Z.~Komargodski, and N.~Seiberg,
  ``{Comments on Chern-Simons Contact Terms in Three Dimensions},''
  \href{http://dx.doi.org/10.1007/JHEP09(2012)091}{{\em JHEP} {\bf 09} (2012)
  091},
\href{http://arxiv.org/abs/1206.5218}{{\tt arXiv:1206.5218 [hep-th]}}.

\bibitem{witten2007}
E.~Witten, ``Three-dimensional gravity revisited,'' {\em arXiv preprint
  arXiv:0706.3359} (2007)  .

\bibitem{hughes2011torsional}
T.~L. Hughes, R.~G. Leigh, and E.~Fradkin, ``Torsional response and
  dissipationless viscosity in topological insulators,'' {\em Phys. Rev. Lett.}
  {\bf 107} (2011) no.~7, 075502.

\bibitem{Hughes:2012vg}
T.~L. Hughes, R.~G. Leigh, and O.~Parrikar, ``{Torsional Anomalies, Hall
  Viscosity, and Bulk-boundary Correspondence in Topological States},''
  \href{http://dx.doi.org/10.1103/PhysRevD.88.025040}{{\em Phys.Rev.} {\bf D88}
  (2013) no.~2, 025040},
\href{http://arxiv.org/abs/1211.6442}{{\tt arXiv:1211.6442 [hep-th]}}.

\bibitem{parrikar2014}
O.~Parrikar, T.~L. Hughes, and R.~G. Leigh, ``Torsion, parity-odd response, and
  anomalies in topological states,'' {\em Phys. Rev. D} {\bf 90} (2014) no.~10,
  105004.

\bibitem{avron1995}
J.~Avron, R.~Seiler, and P.~G. Zograf, ``Viscosity of quantum hall fluids,''
  {\em Phys. Rev. Lett.} {\bf 75} (1995) no.~4, 697.

\bibitem{read2009}
N.~Read, ``Non-abelian adiabatic statistics and hall viscosity in quantum hall
  states and p x+ i p y paired superfluids,'' {\em Phys. Rev. B} {\bf 79}
  (2009) no.~4, 045308.

\bibitem{Solodukhin:1994yz}
S.~N. Solodukhin, ``{The Conical singularity and quantum corrections to entropy
  of black hole},'' \href{http://dx.doi.org/10.1103/PhysRevD.51.609}{{\em Phys.
  Rev.} {\bf D51} (1995)  609--617},
\href{http://arxiv.org/abs/hep-th/9407001}{{\tt arXiv:hep-th/9407001
  [hep-th]}}.

\bibitem{Casini:2011kv}
H.~Casini, M.~Huerta, and R.~C. Myers, ``{Towards a derivation of holographic
  entanglement entropy},''
  \href{http://dx.doi.org/10.1007/JHEP05(2011)036}{{\em JHEP} {\bf 05} (2011)
  036},
\href{http://arxiv.org/abs/1102.0440}{{\tt arXiv:1102.0440 [hep-th]}}.

\bibitem{hasan2010}
M.~Z. Hasan and C.~L. Kane, ``Colloquium: topological insulators,'' {\em Rev.
  Mod. Phys.} {\bf 82} (2010) no.~4, 3045.

\bibitem{Blanco:2013joa}
D.~D. Blanco, H.~Casini, L.-Y. Hung, and R.~C. Myers, ``{Relative Entropy and
  Holography},'' \href{http://dx.doi.org/10.1007/JHEP08(2013)060}{{\em JHEP}
  {\bf 1308} (2013)  060},
\href{http://arxiv.org/abs/1305.3182}{{\tt arXiv:1305.3182 [hep-th]}}.

\bibitem{'tHooft1985727}
G.~'t~Hooft, ``On the quantum structure of a black hole,''
  \href{http://dx.doi.org/http://dx.doi.org/10.1016/0550-3213(85)90418-3}{{\em
  Nuclear Physics B} {\bf 256} (1985)  727 -- 745}.
  \url{http://www.sciencedirect.com/science/article/pii/0550321385904183}.

\bibitem{PhysRevLett.101.010504}
H.~Li and F.~D.~M. Haldane,
  \href{http://dx.doi.org/10.1103/PhysRevLett.101.010504}{``Entanglement
  spectrum as a generalization of entanglement entropy: Identification of
  topological order in non-abelian fractional quantum hall effect states,''{\em
  Phys. Rev. Lett.} {\bf 101} (Jul, 2008)  010504}.
  \url{http://link.aps.org/doi/10.1103/PhysRevLett.101.010504}.

\bibitem{fidkowski2010}
L.~Fidkowski, ``Entanglement spectrum of topological insulators and
  superconductors,'' {\em Phys. Rev. Lett.} {\bf 104} (2010) no.~13, 130502.

\bibitem{turner2010}
A.~M. Turner, Y.~Zhang, and A.~Vishwanath, ``Entanglement and inversion
  symmetry in topological insulators,'' {\em Physs Rev. B} {\bf 82} (2010)
  no.~24, 241102.

\bibitem{sterdyniak2011}
A.~Sterdyniak, N.~Regnault, and B.~Bernevig, ``Extracting excitations from
  model state entanglement,'' {\em Physical review letters} {\bf 106} (2011)
  no.~10, 100405.

\bibitem{PhysRevB.84.195103}
A.~Alexandradinata, T.~L. Hughes, and B.~A. Bernevig,
  \href{http://dx.doi.org/10.1103/PhysRevB.84.195103}{``Trace index and
  spectral flow in the entanglement spectrum of topological insulators,''{\em
  Phys. Rev. B} {\bf 84} (Nov, 2011)  195103}.
  \url{http://link.aps.org/doi/10.1103/PhysRevB.84.195103}.

\bibitem{chandran2011}
A.~Chandran, M.~Hermanns, N.~Regnault, and B.~A. Bernevig, ``Bulk-edge
  correspondence in entanglement spectra,'' {\em Physical Review B} {\bf 84}
  (2011) no.~20, 205136.

\bibitem{hughes2011}
T.~L. Hughes, E.~Prodan, and B.~A. Bernevig, ``Inversion-symmetric topological
  insulators,'' {\em Phys. Rev. B} {\bf 83} (2011) no.~24, 245132.

\bibitem{PhysRevLett.108.196402}
X.-L. Qi, H.~Katsura, and A.~W.~W. Ludwig,
  \href{http://dx.doi.org/10.1103/PhysRevLett.108.196402}{``General
  relationship between the entanglement spectrum and the edge state spectrum of
  topological quantum states,''{\em Phys. Rev. Lett.} {\bf 108} (May, 2012)
  196402}. \url{http://link.aps.org/doi/10.1103/PhysRevLett.108.196402}.

\bibitem{PhysRevB.86.045117}
B.~Swingle and T.~Senthil,
  \href{http://dx.doi.org/10.1103/PhysRevB.86.045117}{``Geometric proof of the
  equality between entanglement and edge spectra,''{\em Phys. Rev. B} {\bf 86}
  (Jul, 2012)  045117}.
  \url{http://link.aps.org/doi/10.1103/PhysRevB.86.045117}.

\bibitem{creutz2001}
M.~Creutz, ``Aspects of chiral symmetry and the lattice,'' {\em Rev. Mod.
  Phys.} {\bf 73} (2001) no.~1, 119.

\bibitem{peschel2004}
I.~Peschel, ``{On the reduced density matrix for a chain of free electrons},''
  {\em {J. Stat. Mech. Theor. Exp.}} {\bf {2004}} ({2004}) no.~{06}, {P06004}.

\bibitem{Guo:2015uqa}
W.-z. Guo and R.-x. Miao, ``{Entropy for gravitational Chern-Simons terms by
  squashed cone method},''
\href{http://arxiv.org/abs/1506.08397}{{\tt arXiv:1506.08397 [hep-th]}}.

\bibitem{Azeyanagi:2015uoa}
T.~Azeyanagi, R.~Loganayagam, and G.~S. Ng, ``{Holographic Entanglement for
  Chern-Simons Terms},''
\href{http://arxiv.org/abs/1507.02298}{{\tt arXiv:1507.02298 [hep-th]}}.

\bibitem{Bak:2003jk}
D.~Bak, M.~Gutperle, and S.~Hirano, ``{A Dilatonic deformation of AdS(5) and
  its field theory dual},''
  \href{http://dx.doi.org/10.1088/1126-6708/2003/05/072}{{\em JHEP} {\bf 05}
  (2003)  072},
\href{http://arxiv.org/abs/hep-th/0304129}{{\tt arXiv:hep-th/0304129
  [hep-th]}}.

\bibitem{Leigh:2008tt}
R.~G. Leigh, N.~N. Hoang, and A.~C. Petkou, ``{Torsion and the Gravity Dual of
  Parity Symmetry Breaking in AdS(4) / CFT(3) Holography},''
  \href{http://dx.doi.org/10.1088/1126-6708/2009/03/033}{{\em JHEP} {\bf 03}
  (2009)  033},
\href{http://arxiv.org/abs/0809.5258}{{\tt arXiv:0809.5258 [hep-th]}}.

\bibitem{Azeyanagi:2007qj}
T.~Azeyanagi, A.~Karch, T.~Takayanagi, and E.~G. Thompson, ``{Holographic
  calculation of boundary entropy},''
  \href{http://dx.doi.org/10.1088/1126-6708/2008/03/054}{{\em JHEP} {\bf 03}
  (2008)  054--054},
\href{http://arxiv.org/abs/0712.1850}{{\tt arXiv:0712.1850 [hep-th]}}.

\bibitem{Bisognano:1976za}
J.~J. Bisognano and E.~H. Wichmann, ``{On the Duality Condition for Quantum
  Fields},''
\href{http://dx.doi.org/10.1063/1.522898}{{\em J. Math. Phys.} {\bf 17} (1976)
  303--321}.

\bibitem{Nieh:1981xk}
H.~T. Nieh and M.~L. Yan, ``{Quantized Dirac Field in Curved Riemann-cartan
  Background. 1. Symmetry Properties, Green's Function},''
\href{http://dx.doi.org/10.1016/0003-4916(82)90186-5}{{\em Annals Phys.} {\bf
  138} (1982)  237}.

\bibitem{EislerPeschel}
V.~Eisler and I.~Peschel, ``Free-fermion entanglement and spheroidal
  functions,'' {\em Journal of Statistical Mechanics: Theory and Experiment}
  {\bf 2013} (2013) no.~04, P04028.
  \url{http://stacks.iop.org/1742-5468/2013/i=04/a=P04028}.

\end{thebibliography}
\providecommand{\href}[2]{#2}\begingroup\raggedright\endgroup

\end{document}